\documentclass[a4paper,twocolumn,11pt,accepted=2021-03-09]{quantumarticle}
\pdfoutput=1
\usepackage[utf8]{inputenc}
\usepackage[english]{babel}
\usepackage[T1]{fontenc}
\usepackage{amsmath}
\usepackage{hyperref}
\usepackage{graphicx}
\usepackage{subfigure}

\usepackage{amssymb}
\usepackage{amsthm}
\usepackage{tikz}
\usepackage{lipsum}

\newcommand{\ket}[1]{| #1 \rangle}

\newcommand{\vt}[1]{{\boldsymbol{#1}}}

\begin{document}

\title{Error mitigation on a near-term quantum photonic device}

\author{Daiqin Su}
\affiliation{Xanadu, Toronto, Ontario, M5G 2C8, Canada}
\email{sudaiqin@gmail.com}
\author{Robert Israel}
\affiliation{Xanadu, Toronto, Ontario, M5G 2C8, Canada}
\author{Kunal Sharma}
\affiliation{Hearne Institute for Theoretical Physics and Department of Physics and Astronomy, Louisiana State University, Baton Rouge, LA USA}
\author{Haoyu Qi}
\affiliation{Xanadu, Toronto, Ontario, M5G 2C8, Canada}
\author{Ish Dhand}
\affiliation{Xanadu, Toronto, Ontario, M5G 2C8, Canada}
\author{Kamil Br\'adler}
\affiliation{Xanadu, Toronto, Ontario, M5G 2C8, Canada}
\maketitle

\begin{abstract}
Photon loss is destructive to the performance of quantum photonic devices and therefore suppressing the effects of photon loss is paramount to photonic quantum technologies. 
We present two schemes to mitigate the effects of photon loss for a Gaussian Boson Sampling device, in particular, to improve the estimation of the sampling probabilities.
Instead of using error correction codes
which are expensive in terms of their hardware resource overhead, our schemes require only a small amount of hardware modifications or even no modification. 
Our loss-suppression techniques rely either on collecting additional measurement data or on classical post-processing once the measurement data is obtained.
We show that with a moderate cost of classical post processing, the effects of photon loss can be significantly suppressed for a certain amount of loss. 
The proposed schemes are thus a key enabler for applications of near-term photonic quantum devices. 
\end{abstract}

\section{Introduction}

Error is the main hindrance for large scale quantum computation. Quantum error correction codes are introduced to correct the errors and allow for 
fault-tolerant quantum computation. However, the conditions for fault-tolerant quantum computation are extremely stringent, requiring very low-error gates and a 
large amount of physical qubits, e.g., about a thousand or more physical qubits are needed to construct a single fault-tolerant logical qubit \cite{fowler2012surface}. Currently, the state of the art technologies 
allow for building quantum devices consisting of about fifty noisy physical qubits, which is very far away from the number required for fault-tolerant computation. 
It is expected that in the intermediate future quantum devices with hundreds of noisy physical qubits are available~\cite{preskill2018quantum}. Efforts have been made to discover 
algorithms that are compatible with these noisy quantum devices and are capable of demonstrating the advantages of quantum computing. Examples include
the Random Circuit Sampling \cite{boixo2018characterizing, aaronson2016complexity, arute2019quantum}, IQP \cite{bremner2011classical, bremner2016average, bremner2017achieving}, Boson Sampling \cite{aaronson2011computational}, Gaussian Boson Sampling (GBS)~\cite{hamilton2017gaussian, rahimi2015what, rahimi2016sufficient}, 
Variational Quantum Eigensolver \cite{peruzzo2014VQE}, and the Quantum Approximate Optimization Algorithm \cite{farhi2014quantum,farhi2016quantum}, etc.

To improve the performance of these near-term quantum devices, several error mitigation techniques have been developed to suppress the noise. One of the promising 
error mitigation techniques exploits extrapolation to approximately estimate the expectation values of some observables of a noise-free circuit by using the measured
expectation values from noisy quantum circuits with various error rates~\cite{Temme2017a, Li2017a}. 
A proof of principle experiment has been performed in a superconducting device and the accuracy
of the variational eigensolver has been significantly improved~\cite{Kandala2019a}. 
Another error mitigation technique that is capable of improving the estimation of expectation values is called quasi-probability decomposition technique, 
in which a noise-free circuit is simulated by a collection of randomly selected noisy circuits following 
a particular quasi-probability distribution~\cite{Temme2017a, Endo2018a}. This technique has also been tested in various experimental platforms~\cite{Songeaaw5686, zhang2020error}. Instead of estimating the expectation values, 
there are error mitigation techniques that can partially recover the error-free quantum states, though more resources are required. One of those examples is to measure
the symmetries of the quantum systems and project it back into the error-free subspace~\cite{Bonet-Monroig2018a, sagastizabal2019experimental, McArdle2019a, bonet2018low}. Another method is to employ variational quantum algorithms to mitigate errors in the state preparation circuits \cite{cerezo2020variational}. 
Moreover, inherent noise resilience of quantum algorithms can lead to error mitigation \cite{mcclean2016theory, sharma2020noise, cincio2020machine}. Other error mitigation techniques have also been developed to correct measurement errors \cite{chen2019detector, geller2020efficient, funcke2020measurement, kwon2020hybrid}, to mitigate decoherence \cite{mcclean2017hybrid}, and for analog quantum simulation \cite{sun2020practical}. Furthermore, learning-based algorithms have also been designed to mitigate errors \cite{strikis2020learning, czarnik2020error, zlokapa2020deep}.

All previously mentioned error mitigation techniques are specifically tailored for quantum systems with a finite Hilbert space dimension.
It is an open question whether these methods are applicable for computing models based on infinite-dimensional systems. 
Here we focus on the GBS device as a promising near-term photonic infinite-dimensional platform.
It is believed that GBS can be used to demonstrate
computational advantages of quantum computer over classical computer.
Various algorithms based on the GBS device have also been proposed, for example, sampling the dense subgraphs~\cite{arrazola2018using}, 
distinguishing non-isomorphic graphs~\cite{bradler2018graph} and quantifying the similarity of graphs~\cite{schuld2020measuring, bradler2019duality}. 
These applications are sensitive to the imperfections of the GBS device, within which photon loss is the main source of errors for a 
photonic implementation. Schemes to mitigate the effect of photon loss are thus paramount for the applications of the GBS devices. 
However, no error mitigation scheme for photonic systems has been developed to date.

In this work, we propose two schemes to mitigate the effect of photon loss in a GBS device. The first scheme exploits the Richardson 
extrapolation technique and is tailored to extrapolate the probability of a photon number pattern or a collection of patterns for a loss-free GBS circuit. To perform the 
extrapolation, one has to vary the loss of the circuit and measure the probabilities of click patterns for every loss value. This requires only a small modification of the 
circuit but one needs to perform several experiments, depending on the number of chosen loss values. 
The second scheme estimates the probability of a click pattern of a loss-free circuit by linearly combining the probabilities of
click patterns with higher total photon numbers in the presence of loss. This requires no modifications of the circuits since one does not need to vary the loss value as the first scheme. Classical post-processing is required to compute the linear combination coefficients.

This paper is organized as follows. In Sec.~\ref{sec:GBS} we briefly introduce the GBS device and some of its applications. In Sec.~\ref{sec:extrapolation} we discuss the 
error mitigation scheme based on the Richardson extrapolation technique. We detail a standard version and an improved version of the extrapolation technique, with the latter has a better performance for large photon loss. 
We discuss the second scheme, the loss cancellation method, in Sec.~\ref{sec:loss-cancellation}. In Sec.~\ref{sec:example},
we test our error mitigation schemes in an eight-mode circuit and compare their performances. Finally, we summarize our discussion in Sec.~\ref{sec:conclusion},
followed by the Appendix.

\section{Overview of GBS device}\label{sec:GBS}

A GBS device consists of three parts: the input state, linear passive interferometer, and photon 
detectors. The input is usually chosen as a product state of pure single-mode Gaussian states, each of which is fully characterized by its displacement and squeezing. 
The linear passive interferometer implements a unitary transformation to the input state and produces a multimode Gaussian state in the output. In each output mode
a photon detector counts the number of photons. 

Consider a GBS device with $M$ input/output modes. The output multimode Gaussian state is characterized by a $2M$-component vector ${\boldsymbol d}$ and 
a $2M \times 2M$ covariance matrix $\sigma$~\cite{weedbrook2018gaussian}. Assume that $\vt{n} = [n_1, n_2, \dots, n_M]$
represents a certain measurement pattern of photons (a click pattern), 
where $n_j$ is the detected photon number in the $j$-th mode. In the case of no displacement, namely $\vt{d} = \vt{0}$, 
the probability of detecting a click pattern $\vt{n}$ is given by~\cite{hamilton2017gaussian}
\begin{align}
\label{eq:ProbabilityGeneral}
P(\vt{n}) 
&=& \frac{1}{\vt{n}! \, \sqrt{\text{det} \, \sigma_Q}} \prod_{k=1}^M \bigg( \frac{\partial^2}{\partial \alpha_k \partial \alpha_k^* }\bigg)^{n_k} \nonumber\\
&& \times
\exp\bigg(\frac{1}{2} \vt{\alpha}_v^{\top} A \vt{\alpha}_v \bigg) \bigg|_{\vt{\alpha}_v = \vt{0}},
\end{align}
where $\vt{n} ! = n_1 ! n_2 ! \cdots n_M !$, $\sigma_{Q} = \sigma + \mathbb{I}_{2M}/2$, 
$\vt{\alpha}_v = (\alpha_1, \cdots, \alpha_M, \alpha_1^*, \cdots, \alpha_M^*)^{\top}$, and the matrix $A$ is given by
\begin{eqnarray}
\label{eq:Amatrix}
A = X_{2M} \big( \mathbb{I}_{2M} - \sigma_Q^{-1} \big), ~ \text{with} ~
X_{2M} = 
	\begin{pmatrix}
	0 & \mathbb{I}_M \\
	\mathbb{I}_M & 0
	\end{pmatrix}. \nonumber
\end{eqnarray}
The expression for probability in Eq.~\eqref{eq:ProbabilityGeneral} is valid for both pure and mixed Gaussian states. 

It is believed that a GBS device with a sufficiently large number of modes and input photons can generate a photon number probability distribution that is hard to sample from using
a classical computer~\cite{hamilton2017gaussian}. The GBS device can also be used to solve graph-related problems by mapping the adjacency matrix 
of a graph to the covariance matrix of the GBS output state~\cite{bradler2018gaussian}. The properties of the graph are thus encoded into the photon number probability distribution. 
Such applications include sampling the dense subgraphs~\cite{arrazola2018using}, distinguishing non-isomorphic graph~\cite{bradler2018graph,bradler2019duality}, 
quantifying similarity of two graphs~\cite{schuld2020measuring}, and other potential graph-related applications to be discovered. 

A realistic GBS device is noisy, suffering from experimental imperfections like the photon loss, 
thus limiting its computational power~\cite{qi2020regimes}. Given that fault tolerant error correction will not be accessible in the near future, 
it is paramount to develop some error mitigation techniques with least amount of hardware modifications to suppress the effects of photon loss and improve 
the performance of the GBS device. In the following sections, we focus on algorithms that involve estimating sampling probabilities of a click 
pattern or a collection of click patterns, and propose two schemes to improve the estimation of the probability for a loss-free GBS device.

\section{Suppress photon loss via extrapolation}\label{sec:extrapolation}

The first method of suppressing errors in GBS is inspired from the 
extrapolation procedure introduced in Ref.~\cite{Temme2017a}.
Similar to Ref.~\cite{Temme2017a}, our method relies on performing multiple measurements and extrapolating to obtain the desired outcome.
However, while Ref.~\cite{Temme2017a} focuses on decreasing an overall multiplicative factor in the system Hamiltonian thus leading to an effectively higher error rate, we exploit the possibility of directly changing the error rate in photonic systems.
Another novelty of our work is an improved method that provides enhanced performance for GBS probabilities by removing the poles of the extrapolation function before extrapolating as described in Sec.~\ref{Sec:SecondExtrapolation}.

\subsection{General formalism}\label{subsec:GF}

Consider a quantum circuit with an ideal input state and gates, the density operator of the output state before detection is $\hat{\rho}_0$. The expectation value of an observable
$\hat O$ is $O_0 = \text{tr}(\hat O \hat{\rho}_0)$, where ``tr" represents the trace. If the quantum circuit is imperfect, for example, the gate has an error rate characterized by 
a small parameter $\epsilon$, then the output density operator becomes $\hat{\rho}_{\epsilon}$ and the expectation value becomes 
$O({\epsilon}) = \text{tr}(\hat O \hat{\rho}_{\epsilon})$.
Both $\hat{\rho}_{\epsilon}$ and $O({\epsilon})$ deviate from $\hat{\rho}_0$ and $O_0$, respectively, but approach to them in the limit of $\epsilon \rightarrow 0$. 
We thus can perform a series expansion of $\hat{\rho}_{\epsilon}$ as
\begin{eqnarray}
\hat{\rho}_{\epsilon} = \hat{\rho}_0 + \sum_{k=1}^{\infty} \hat{\rho}_k \epsilon^k,
\end{eqnarray}
and similarly for $O({\epsilon})$,
\begin{eqnarray}
O({\epsilon})= O_0 + \sum_{k=1}^{\infty} O_k \epsilon^k,
\end{eqnarray}
where $\hat{\rho}_k$ are operators and $O_k$ are real numbers. 

Suppose one can vary the error rate in a controllable way, in particular, to increase the error rate on purpose. Denote the error rates as $\epsilon_j = \epsilon c_j$, with $j = 0, 1, 2, \cdots, m$,
and we choose $c_0 =1$ and $c_j >1$ for $j \ne 0$. For each error rate $\epsilon_j$, one performs the experiment and measures the expectation value $O(\epsilon_j)$. 
By linearly combining $(m+1)$ measured expectation values,  one arrives at an estimation of the expectation value $O_0$ as
\begin{eqnarray}
\tilde{O}(\epsilon) = \sum_{j=0}^m \gamma_j O(\epsilon_j) = \sum_{j=0}^m \gamma_j O( \epsilon c_j ). 
\end{eqnarray}
The coefficients $\gamma_j$ are appropriately chosen such that $O_k$ for $k=1, 2, \cdots, m$ are cancelled, giving 
\begin{eqnarray}
\tilde{O}(\epsilon) = O_0 + \mathcal{O} (\epsilon^{m+1}),
\end{eqnarray}
a better estimation of the loss-free expectation value when $\epsilon$ is small. Taking this requirement into account, 
we find that  $\gamma_j$ satisfy a linear system of equations
\begin{align}
\label{eq:ConstraintGamma}
\sum_{j=0}^m \gamma_j = 1, ~~ \sum_{j=0}^m \gamma_j c_j^k = 0, ~~~ k = 1, 2, \cdots, m, 
\end{align}
and the solution can be found as
\begin{eqnarray}\label{eq:gamma}
\gamma_j = (-1)^m \prod_{k \ne j}^m \frac{c_k}{c_j - c_k}. 
\end{eqnarray}
This extrapolation technique has been applied to qubit systems.

\subsection{Extrapolate sampling probability}

We now apply the above technique to suppress the photon loss of a GBS device and obtain a better estimation of the sampling probability. To illustrate the method, we first 
consider the uniform loss case, namely, the photon loss in each mode is the same and is characterized by a single parameter $\epsilon$. This is a good approximation to a 
realistic linear interferometer if it is implemented using the Clements' decomposition~\cite{Clements2016} (see Appendix~\ref{ap:ULA} for more details). 

Assume that the covariance matrix of the output state without photon loss is $\sigma_0$ and that with photon loss is $\sigma_{\epsilon}$. They are related via the action of a 
lossy bosonic channel, 
\begin{eqnarray}
\sigma_{\epsilon} = (1 - \epsilon) \sigma_0 + \frac{1}{2} \epsilon \, \mathbb{I}_{2M},
\end{eqnarray}
where $\mathbb{I}_{2M}$ is the identity matrix. 
We thus have
\begin{eqnarray}\label{eq:sigmaQloss}
\sigma_Q(\epsilon) &=& \sigma_{\epsilon} + \frac{1}{2}\, \mathbb{I}_{2M} 
\nonumber\\
&=& \sigma_0 + \frac{1}{2}\, \mathbb{I}_{2M} - \frac{1}{2}\, \epsilon \, (2 \sigma_0 - \mathbb{I}_{2M}), 
\end{eqnarray}
and 
\begin{eqnarray}\label{eq:InverseSQ}
[\sigma_Q(\epsilon)]^{-1} 
&=& 2(2 \sigma_0 + \mathbb{I}_{2M})^{-1} \bigg[ \mathbb{I}_{2M} + \nonumber\\
&&  \sum_{k=1}^{\infty} (2 \sigma_0 - \mathbb{I}_{2M})^k (2 \sigma_0 + \mathbb{I}_{2M})^{-k} \, \epsilon^k \bigg]. \nonumber
\end{eqnarray}
By substituting this into Eq.~\eqref{eq:Amatrix}, we obtain a series expansion of the $A$ matrix as
\begin{eqnarray}\label{eq:Aexpansion}
A(\epsilon) = A_0 + \sum_{k=1}^{\infty} A_k \epsilon^k,
\end{eqnarray}
where $A_0$ is the matrix corresponding to no photon loss state and the coefficients $A_k$ is given by
\begin{align}\label{eq:ExpansionAk}
A_k 
&=& - 2 X_{2M} (2 \sigma_0 - \mathbb{I}_{2M})^k (2 \sigma_0 + \mathbb{I}_{2M})^{-k-1}. 
\end{align}
The probability, $P(\vt{n}; \epsilon)$, of a click pattern $\vt{n}$ with photon loss $\epsilon$ can be obtained by replacing the $A$ matrix and 
$\sigma_Q$ in Eq.~\eqref{eq:ProbabilityGeneral} by $A(\epsilon)$ and $\sigma_Q(\epsilon)$, respectively. Since both $A(\epsilon)$ and $\sigma_Q(\epsilon)$
have series expansions with respect to $\epsilon$, we thus can find a series expansion for $P(\vt{n}; \epsilon)$ as 
\begin{eqnarray}\label{eq:ProbabilitySeries}
P(\vt{n}; \epsilon) = P_0(\vt{n}) + \sum_{k=1}^{\infty} P_k(\vt{n}) \epsilon^k,
\end{eqnarray}
where $P_0(\vt{n})$ corresponds to the sampling probability without photon loss and is the quantity that we want to estimate,
and $P_k(\vt{n})$ is a complicated expression and its explicit form is not relevant here. 

Now we apply the general formalism developed in Sec.~\ref{subsec:GF} to derive a better estimation of $P_0(\vt{n})$. Assume that one can choose different 
loss values, $\epsilon_j$, and estimate the corresponding probability, $P(\vt{n}; \epsilon_j)$, in the experiment. Similarly, define $\epsilon_j = \epsilon c_j$, with $j = 0, 1, 2, \cdots, m$,
and we choose $c_0 =1$ and $c_j >1$ for $j \ne 0$. Then we get a better estimation of the probability $P_0(\vt{n})$ by defining
\begin{align}\label{eq:GoodEstimate}
\tilde P(\vt{n}, \epsilon) = \sum_{j=0}^m \gamma_j P(\vt{n}; \epsilon_j) = \sum_{j=0}^m \gamma_j P(\vt{n}; \epsilon c_j ),
\end{align}
which further simplifies to
\begin{eqnarray}
\tilde{P}(\vt{n}; \epsilon) = P_0(\vt{n}) + \mathcal{O}(\epsilon^{m+1}), 
\end{eqnarray}
for $\gamma_j$ as in Eq.~\eqref{eq:gamma} and for small values of $\epsilon$. 
Different values of loss are indeed possible in the experiment: specifically, loss can be programmably increased in experiment using $M$ additional tunable beam splitters that have one output port discarded. Experimentally, a tunable beam splitter is implemented by a Mach-Zehnder interferometer consisting of
two static 50:50 beam splitters and two phase shifters ~\cite{Reck1994} (see also Appendix~\ref{ap:ULA} for more details). 
Its transmission coefficient can be tuned by varying the phases, e.g., via changing the temperature of the device~\cite{Jacques:19}.
Suppose the transmission coefficient of each additional beam splitter is $\eta_{\rm ad}$, then the overall transmission of the circuit is modified to be $(1-\epsilon)\eta_{\rm ad}$. 
This implies the loss value is changed to $\epsilon^{\prime} = 1- (1-\epsilon) \eta_{\rm ad} = \epsilon [ \eta_{\rm ad} + (1- \eta_{\rm ad})/\epsilon]$. One can appropriately choose
the value of $\eta_{\rm ad}$ to attain a target loss value $\epsilon_j$. 
A potential challenge is that the detection probability of the needed click patterns decreases and thus it requires a lot of data collection.

\subsection{A two-mode squeezed vacuum example}\label{sec:TMSVextrapolation}

\begin{figure}
\includegraphics[width=0.96\columnwidth]{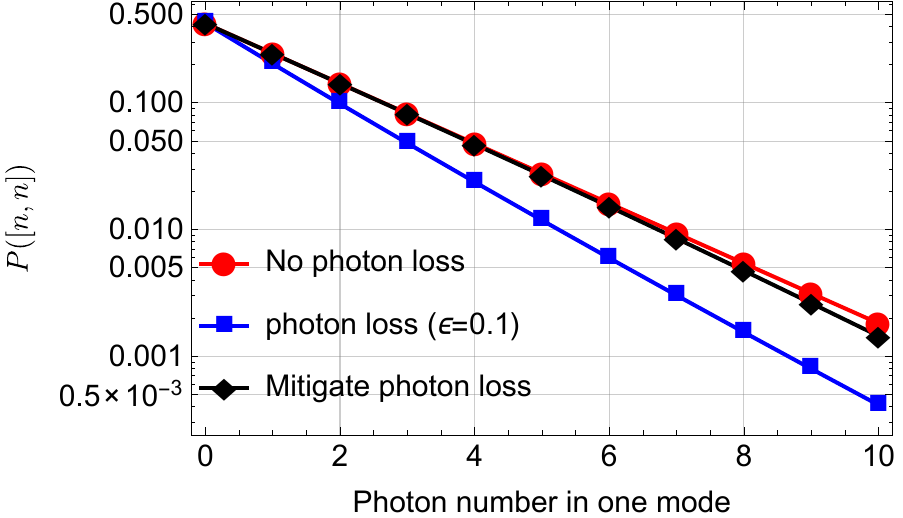}
\caption{ Error mitigation to a lossy TMSV state with squeezing parameter $r = 1.0$. The red circles represent probabilities for 
a pure TMSV state, the blue squares represent probabilities for a lossy TMSV state and the black rhombuses represent the extrapolated probabilities with 
photon loss $\epsilon = 0.1$.  }
\label{fig:TMSV-error-mitigation}
\end{figure}

To showcase the extrapolation technique, we consider a simple example: to mitigate photon loss in a two-mode squeezed vacuum  (TMSV) state. 
A TMSV state is defined as
\begin{eqnarray}\label{eq:TMSV}
\ket{\chi}_{\text{TMSV}} = \sqrt{1 - \chi^2} \sum_{n=0}^{\infty} \chi^n \ket{n} \ket{n},
\end{eqnarray}
where $\chi = \tanh r$ and $r$ is the squeezing parameter~\cite{Serafini2017a}. When the TMSV state is detected by two photon-number-resolving (PNR) detectors, 
the only possible detected photon number patterns are $\vt{n} = [n, n]$ and the probability is 
\begin{align}
P_0([n, n]) = (1 - \chi^2) \chi^{2n} = \frac{(\tanh r)^{2n}}{\cosh^2 r}. 
\end{align}
The measurement probability $P_0(n, n)$ for $r = 1.0$ is plotted in Fig.~\ref{fig:TMSV-error-mitigation} (red circle). 
Now a pure lossy channel with transmissivity $\eta = 1 - \epsilon$ is added to each output mode of the TMSV state, resulting in a mixed two-mode Gaussian state. 
The photon number distribution will be modified, which we denote as $P(n, m; \epsilon)$. We then plot $P(n, n; \epsilon)$ for $r = 1.0$ and $\epsilon  = 0.1$
in Fig.~\ref{fig:TMSV-error-mitigation} (blue square), which has a big deviation from the no photon loss case.

Now we apply the extrapolation technique to mitigate the photon loss. We choose $m=4$ and $\vt{c} = (1.0, 1.2, 1.4, 1.6, 1.8)$. 
From Eq.~\eqref{eq:gamma}, we find $\vt{\gamma} = (126, -420, 540, -315, 70)$. The new approximation to the ideal case can be evaluated
directly from Eq.~\eqref{eq:GoodEstimate}. We plot $\tilde P(n, n; \epsilon)$ for $r = 1.0$ and $\epsilon  = 0.1$ in Fig.~\ref{fig:TMSV-error-mitigation} (black rhombus).
We see that $\tilde P(n, n; \epsilon)$ is much closer to $P_0(n, n)$ as compared to $P(n, n; \epsilon)$, showing that the extrapolation technique significantly 
mitigate the effect of photon loss. After the error mitigation, the effective photon loss is approximately $\epsilon^{\prime} \approx 0.01$. 

\subsection{Improved extrapolation of sampling probability}
\label{Sec:SecondExtrapolation}

We have shown that the extrapolation technique works very well for low photon loss. However, when the amount of photon loss and 
input squeezing increase, the extrapolated result becomes less and less accurate. We find that by slightly modifying the previous extrapolation procedure, one can obtain a better 
estimation of the probability for large squeezing and relatively high photon loss. 

By using the expression of $\sigma_Q(\epsilon)$ in Eq.~\eqref{eq:sigmaQloss}, one can rewrite the $A$ matrix with photon loss as
\begin{align}
A(\epsilon) 
&= (1-\epsilon) X_{2M} (2 \sigma_0 - \mathbb{I}_{2M}) \nonumber\\
& \times [(2 \sigma_0 + \mathbb{I}_{2M}) - (2 \sigma_0 - \mathbb{I}_{2M}) \epsilon]^{-1}.
\end{align}
There exists a unitary matrix $U$ diagonalizes the the covariance matrix $\sigma_0$ as 
\begin{eqnarray}\label{eq:CMdiagonalize}
\sigma_0 = \frac{1}{2} U \bigoplus_{k=1}^{M}
	\begin{pmatrix}
	e^{2 r_k} & 0 \\
	0 & e^{-2 r_k}
	\end{pmatrix}
U^{\dag},
\end{eqnarray}
where we assume $\sigma_0$ is a pure-state covariance matrix and $r_k$ are the squeezing parameters of the input squeezed vacuum states. 
Here $U$ represents the transformation of the linear interferometer, and is independent of the input squeezing and the photon loss. Then $A(\epsilon)$ can be rewritten as
\begin{eqnarray}\label{eq:Aepsilon}
&& A(\epsilon) 
\nonumber\\
&=& 
(1-\epsilon) X_{2M} U \bigoplus_{k=1}^{M}
	\begin{pmatrix}
	\frac{\tanh r_k}{1-\epsilon \tanh r_k} & 0 \\
	0 & -\frac{\tanh r_k}{1+\epsilon \tanh r_k}
	\end{pmatrix}
U^{\dag}
\nonumber\\
&\equiv&
\frac{1-\epsilon}{\mathcal{P}(\epsilon, \tilde{\vt{r}})} R(\epsilon, \vt{r}),
\end{eqnarray}
where $\vt{r} = (r_1, r_2, \cdots, r_M)^{\top}$ and $\tilde{\vt{r}} = (\tilde{r}_1, \tilde{r}_2, \cdots, \tilde{r}_{N_{\lambda}})^{\top}$, with $\{\tilde{r}_j\}_{j\in\{1,\dots, N_{\lambda}\}  }$ denotes a set of nonzero different squeezing parameters in the set $\{r_k\}_{k\in \{1, \dots, M\}  }$ 
, and we defined
\begin{eqnarray}
\mathcal{P}(\epsilon, \tilde{\vt{r}}) &=& \prod_{j=1}^{N_{\lambda}} (1-\epsilon^2 \tanh^2 \tilde{r}_j),\nonumber \\ 
R(\epsilon,\vt{r}) &=&
\mathcal{P}(\epsilon, \tilde{\vt{r}}) X_{2M} 
\nonumber\\
&&
\times
U \bigoplus_{k=1}^{M}
\begin{pmatrix}
\frac{\tanh r_k}{1-\epsilon \tanh r_k} & 0 \\
0 & -\frac{\tanh r_k}{1+\epsilon \tanh r_k} 
\end{pmatrix}
U^{\dag}. \nonumber
\end{eqnarray}
\noindent It is evident that $\mathcal{P}(\epsilon, \tilde{\vt{r}})$ is a polynomial of order $2 N_{\lambda}$, and $\mathcal{P}(\epsilon, \tilde{\vt{r}})/(1 \pm \epsilon \tanh r_k)$ is a polynomial of 
order $2 N_{\lambda}-1$ if $r_k \neq 0$ and of order $2 N_{\lambda}$ if $r_k = 0$. Since $X_{2M}$ is a constant matrix and $S$ (represents the linear interferometer) 
is independent of photon loss $\epsilon$ and input squeezing $r_j$, so every entry of the matrix $R(\epsilon,\vt{r})$ is a polynomial of $\epsilon$ of order at most $2N_{\lambda}$ (when
at least one of the $r_j$ is zero), or $2N_{\lambda}-1$ (when all $r_j$ are not zero). 

From Eq. \eqref{eq:InverseSQ} we find 
\begin{eqnarray}
\text{det} \big\{ [\sigma_{Q}(\epsilon)]^{-1/2} \big\} &=& \prod_{k=1}^M \frac{1}{\cosh r_k} \frac{1}{\sqrt{1 - \epsilon^2 \tanh^2 r_k}} \nonumber\\
&=& \bigg( \prod_{k=1}^M \frac{1}{\cosh r_k} \bigg) \frac{1}{\mathcal{Q} (\epsilon, \vt{r})},
\end{eqnarray}
where $\mathcal{Q} (\epsilon, \vt{r}) \equiv \prod_{k=1}^M \sqrt{1 - \epsilon^2 \tanh^2 r_k}$. 
By substituting $A(\epsilon)$ and $\sigma_Q(\epsilon)$ into Eq.~\eqref{eq:ProbabilityGeneral}, we find
the probability of measuring a click pattern $\vt{n}$ in the presence of photon loss can be written as
\begin{align}\label{eq:PoleSeparate}
P(\vt{n}; \epsilon) = \frac{1}{\vt{n}!} \bigg( \prod_{k=1}^M \frac{1}{\cosh r_k} \bigg) 
\frac{(1-\epsilon)^N}{\mathcal{Q} (\epsilon, \vt{r}) \mathcal{P}^N(\epsilon, \tilde{\vt{r}})} \, \mathbb{P}(\epsilon), 
\nonumber\\
\end{align}
where $N=\sum_{j=1}^M n_j$ is the total detected photon number,
$\mathbb{P}(\epsilon)$ is a polynomial of $\epsilon$ of order at most $2 N N_{\lambda}$ (when at least one of the $r_j$ is zero), or $N(2N_{\lambda}-1)$ 
(when all $r_j$ are not zero).

Notice that $\mathcal{Q} (\epsilon, \vt{r}) \rightarrow 0$ and $\mathcal{P} (\epsilon, \tilde{\vt{r}}) \rightarrow 0$ when $\epsilon \rightarrow 1$ and $r_k \rightarrow \infty$, namely,
both $\mathcal{Q} (\epsilon, \vt{r})$ and $\mathcal{P} (\epsilon, \tilde{\vt{r}})$ are close to zero in the large squeezing and photon loss regime. 
Therefore, $\mathcal{Q} (\epsilon, \vt{r})$ and $\mathcal{P} (\epsilon, \tilde{\vt{r}})$ contribute to the ``singular" part of the probability expression $P(\vt{n}; \epsilon)$, and the 
corresponding poles are given by $\epsilon = \coth r_k > 1$. The presence of these poles limits the accuracy of the first extrapolation procedure. To obtain a better extrapolation
accuracy, we can remove these poles by multiplying $P(\vt{n}; \epsilon)$ with $\mathcal{Q} (\epsilon, \vt{r}) \mathcal{P}^N(\epsilon, \tilde{\vt{r}})$, and then perform the linear combination.
That is to say, the new estimation for the probability without photon loss is 
\begin{align}\label{eq:BetterEstimate}
\bar P(\vt{n}, \epsilon) 
= \sum_{j=0}^m \gamma_j P(\vt{n}; c_j \epsilon) \mathcal{Q} (c_j \epsilon, \vt{r}) \mathcal{P}^N(c_j \epsilon, \tilde{\vt{r}}). 
\nonumber\\
\end{align}
Notice that $\mathcal{Q} (\epsilon, \vt{r}) \rightarrow 1$ and $\mathcal{P} (\epsilon, \tilde{\vt{r}}) \rightarrow 1$ when $\epsilon \rightarrow 0$, so the estimation 
$\bar P(\vt{n}, \epsilon)$ approaches to $P_0(\vt{n})$ in the limit of $\epsilon \rightarrow 0$.

\begin{table*}[ht]
\caption{ Comparing the performance of the improved extrapolation and the normal extrapolation technique for a 
TMSV state. The first column is the click pattern and the second column gives the probability for a pure TMSV state. The 
third and fourth main columns compare results for loss $\epsilon = 0.2$ and $\epsilon =0.5$, respectively. The symbol ``extrap." stands for
extrapolation, and ``imp. extrap." stands for improved extrapolation. } 
\centering 
\resizebox{0.9\textwidth}{!}{\begin{minipage}{\textwidth}
\centering
\begin{tabular}{ c | c |cc|cc}
\hline \hline
  & & $\epsilon=0.2$ & & $\epsilon=0.5$ & \\
  ~$[n, n]$ ~& exact & extrap. & ~ imp. extrap. & extrap. & ~ imp. extrap. \\
    \hline
~$[0,0]$ ~& 0.4200 & 0.4202 & 0.4200 &  0.8406 & 0.4200 \\
~$[1,1]$ ~& 0.2436 & 0.2429 & 0.2436 &  0.3125 & 0.2436  \\
~$[2,2]$ ~& 0.1413 & 0.1387 & 0.1400 &  0.1597 & 0.1140 \\
~$[3,3]$ ~& 0.0820 & 0.0770 & 0.0781 &  0.0308 & 0.0701 \\
~$[4,4]$ ~& 0.0475 & 0.0415 & 0.0421 &  0.0128 & 0.0317 \\
~$[5,5]$ ~& 0.0276 & 0.0218 & 0.0222 &  0.0102 & 0.0115 \\
~$[6,6]$ ~& 0.0160 & 0.0114 & 0.0116 &  0.0068 & 0.0037 \\
\hline \hline
\end{tabular}
\label{tab:ImprovedExtrap}
\end{minipage}}
\end{table*}

To showcase the performance of the improved extrapolation technique, we consider mitigating the photon loss of a TMSV state as in Sec.~\ref{sec:TMSVextrapolation}.
Table~\ref{tab:ImprovedExtrap} compares the results of extrapolation and improved extrapolation for low photon loss $\epsilon = 0.2$ and relatively high photon loss
$\epsilon = 0.5$ cases. We see that for low photon loss, they both give very good approximations to the exact sampling probabilities. While for high photon loss,
the improved extrapolation gives better results for click patterns with low total photon number. In particular, the probabilities for click patterns $[0,0]$ and $[1,1]$ can 
always be exactly extrapolated in the improved extrapolation. This can be understood as follows. After removing the poles, the right hand side of Eq.~\eqref{eq:PoleSeparate} is 
simply a polynomial, so in principle the probability $P_0(\vt{n})$ can be extrapolated exactly if sufficient loss values are chosen, namely, $m$ is at least the same as the order of the 
polynomial. For the TMSV state example, the two input squeezing parameters are the same, so $N_\lambda = 1$. By taking into account the factor $(1-\epsilon)^N$, the 
right hand side of Eq.~\eqref{eq:PoleSeparate} is a polynomial with order $2N$ after removing the poles. Therefore, by taking $m=4$, one can exactly extrapolate probabilities
of click patterns with total photon number less than three. 

From the experiment perspective, removing the poles is only possible when one can fully control the photon loss and know the input squeezing parameters accurately. 
This requires a good calibration of the GBS device in a priori.

\subsection{Extrapolation precision analysis}

We have showed that the extrapolation technique works quite well by simply choosing several loss values, e.g., $m=4$. 
Better results can be obtained by increasing the number of loss values, namely, to increase the number of experiments. 
For improved extrapolation technique, one can in principle extrapolate the exact value of the sampling probability by increasing the number 
of loss values. However, we show that this is challenging in practice. Specifically, the required measurement accuracy should increase exponentially in order to get 
a good extrapolated probability when the number of loss values increases, resulting in exponential increase of running time for the experiment.

To estimate some quantities, like the sampling probability, there is always an uncertainty due to a limited number of samples, or due to the experimental 
imperfections. We now consider how the statistic uncertainty and experimental imperfections affect the extrapolation results. 
From Eq.~\eqref{eq:GoodEstimate} we
can see that the uncertainty in $P(\vt{n}; c_j \epsilon)$ results in uncertainty in $\tilde P(\vt{n}; \epsilon)$. Assume that $\widehat{P}(\vt{n}; c_j \epsilon)$ is an estimator of the 
sampling probability of click pattern $\vt{n}$ with photon loss $c_j \epsilon$ and  $P(\vt{n}; c_j \epsilon)$ is considered as its mean value, then
\begin{eqnarray}\label{eq:uncertainty-in-p}
\widehat{P}(\vt{n}; c_j \epsilon) = P(\vt{n}; c_j \epsilon)  (1 + \widehat X_j),
\end{eqnarray}
where $\widehat X_j$ is a random variable with zero mean and its variance characterizes
the relative uncertainty of the measured probability.
Here we assume that the precision analysis is performed for multiplicative precision, where the variance of $\widehat{X}_j$ does not depend on the 
pattern $\vt{n}$ of photons. A more useful analysis would model the actual error in an experiment and would need to account for inaccuracies in the modelling of the experiment. Such an analysis is left for future work.

From Eq.~\eqref{eq:uncertainty-in-p}, the relative fluctuation of the extrapolated probability $\tilde P (\vt{n}; \epsilon)$ is
\begin{align}
\widehat Y(\vt{n}; \epsilon) &\equiv \frac{\widehat{\tilde P } (\vt{n}; \epsilon) - \tilde P(\vt{n}; \epsilon) }{ \tilde P(\vt{n}; \epsilon) } 
\nonumber\\
& = \frac{\sum_{j=0}^m \gamma_j P(\vt{n}; c_j \epsilon) \widehat X_j}{\sum_{j=0}^m \gamma_j  P(\vt{n}; c_j \epsilon)},
\end{align}
where $\widehat{\tilde P } (\vt{n}; \epsilon)$ is obtained by replacing $P(\vt{n}; c_j \epsilon)$ in Eq.~\eqref{eq:GoodEstimate} by $\widehat{P}(\vt{n}; c_j \epsilon)$. 
It is straightforward to show that the variance of $\widehat Y(\vt{n}; \epsilon)$ is given by
\begin{eqnarray}
\text{Var}( \widehat Y)  = \frac{\sum_{j=0}^m \gamma_j^2  P^2 (\vt{n}; c_j \epsilon) V_j }{\big[\sum_{j=0}^m \gamma_j P(\vt{n}; c_j \epsilon) \big]^2},
\end{eqnarray}
where $V_j$ is the variance of $\widehat X_j$ and we have assumed that $\widehat X_j$ are independent random variables. 
Denote the minimum variance of $\{ \widehat X_j \}_{j=0}^m$ as $V_{\rm min}$ and the minimum nonzero $P (\vt{n}; c_j \epsilon)$ as $P_{\rm min}$; and the corresponding maximums as $V_{\rm max}$  and $P_{\rm max}$. 
Then we can derive a lower bound and an upper bound for $\text{Var}( \widehat Y)$,
\begin{align}\label{eq:Variance}
\text{Var}( \widehat Y) &\ge \frac{  P_{\rm min}^2  V_{\rm min} }{\big[\sum_{j=0}^m \gamma_j P(\vt{n}; c_j \epsilon) \big]^2} \sum_{j=0}^m \gamma_j^2 
\nonumber\\
&= C_1 \, \Gamma_2 V_{\rm min},
\nonumber\\
\text{Var}( \widehat Y) &\le \frac{  P_{\rm max}^2  V_{\rm max} }{\big[\sum_{j=0}^m \gamma_j P(\vt{n}; c_j \epsilon) \big]^2} \sum_{j=0}^m \gamma_j^2 
\nonumber\\
&= C_2 \, \Gamma_2 V_{\rm max},
\end{align}
where $C_1$ and $C_2$ are approximately constants because $\sum_{j=0}^m \gamma_j P(\vt{n}; c_j \epsilon)$ approaches $P_0(n)$, and $P_{\min} (P_{\max})$ becomes constant for large values of $m$. Moreover, $\Gamma_2 = \sum_{j=0}^m \gamma_j^2$ increases exponentially
as the number of loss values, $m$, increases, which follows from Eq. \eqref{eq:gamma}. Therefore, $\text{Var}( \widehat Y)$ increases exponentially 
if $V_{\rm min}$ is fixed, which implies that it becomes exponentially hard to estimate the loss-free sampling probability.

In practice, the strategy is to choose a number of loss values such that a sufficient good approximation to $P_0(\vt{n})$ is obtained while it is still tractable to measure
the probabilities for various photon loss. As an example, we discuss the effect of uncertainty of estimating the probability of a certain click pattern. 
Estimating a probability can be achieved by sampling a
process many times and counting the number of success event. Assume that $N_{\text{succ}}$ and $N$ are the number of success tries and the total
number of tries, respectively. The probability estimated by $\widehat p = N_{\text{succ}}/N$ is a random variable and its variance is given by
\begin{eqnarray}
\text{Var}(\widehat p) = \frac{p(1-p)}{N},
\end{eqnarray}
where $p$ is the mean value of $\widehat p$. The variance of the relative error of $\widehat p$ is 
\begin{eqnarray}
\text{Var} \bigg(\frac{\widehat p}{p} \bigg) = \frac{1-p}{N p}. 
\end{eqnarray}
If this variance is treated as $V_{\rm max}$, then it is straightforward to find the required total number of tries,
\begin{align}
N = \frac{1-p}{p} \, \bigg[ \text{Var} \bigg(\frac{\widehat p}{p} \bigg) \bigg]^{-1} \le \frac{1-p}{p} \frac{ C_2 \Gamma_2 }{\text{Var}( \widehat Y)}.
\end{align}
This gives an estimation of the required total number of tries in order to achieve a certain accuracy of the extrapolated probability, and also the required running time
to collect data if the sampling rate is known. 

\begin{figure}[t]
\includegraphics[width=0.9\columnwidth]{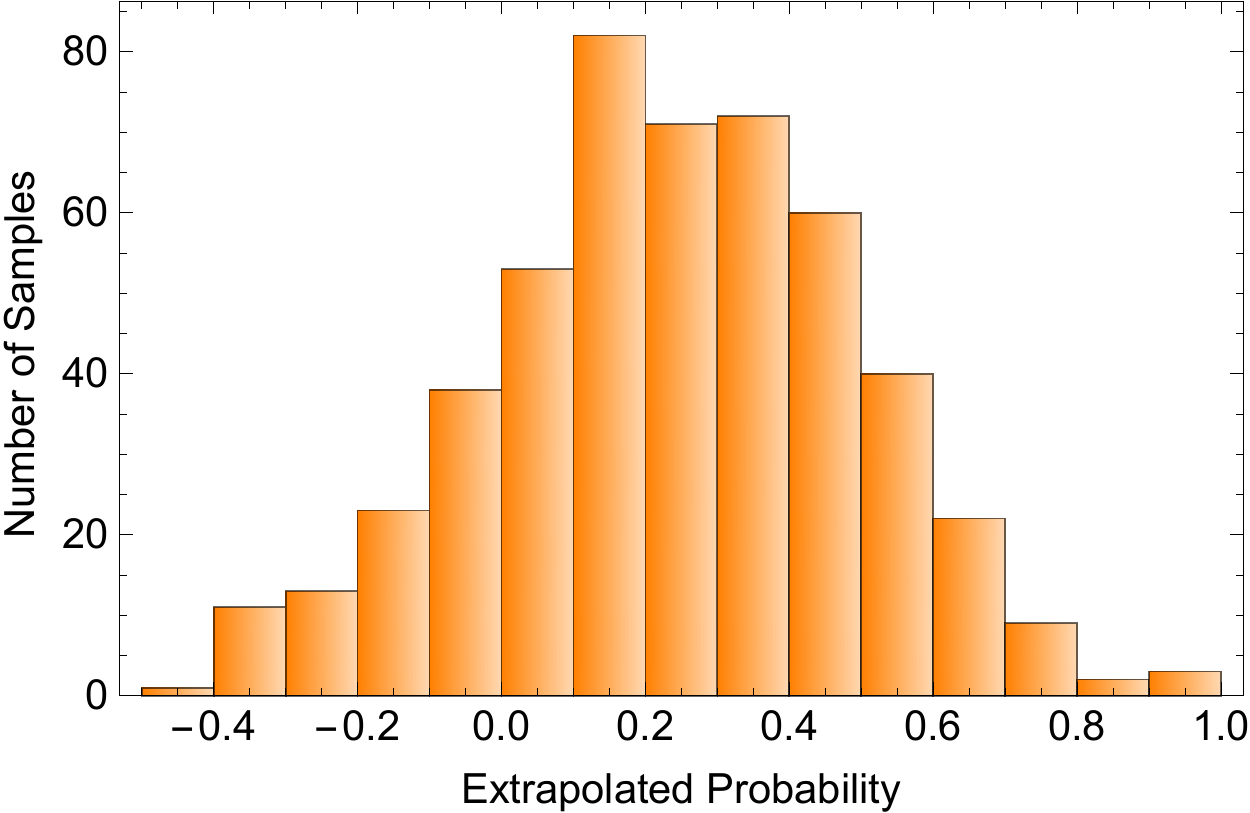}
\caption{ Distribution of extrapolated probability for click pattern $[1, 1]$ due to the uncertainty of probability measurement. 
Here 500 samples of size $N=10^5$ are collected. The mean and standard deviation of the extrapolated probability are 0.2373 and 0.2531, respectively. }
\label{fig:ProbFluctuation}
\end{figure}

\begin{figure}[t]
\includegraphics[width=0.9\columnwidth]{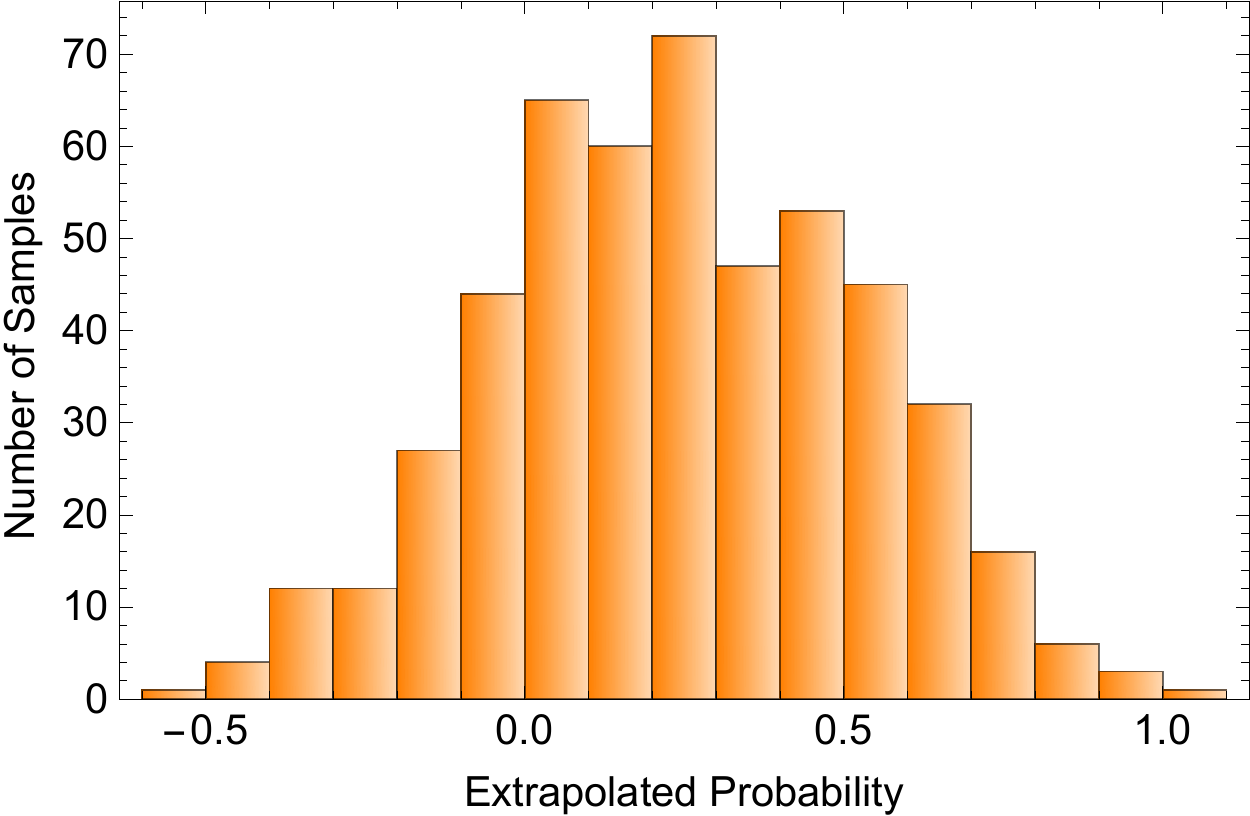}
\caption{ Distribution of extrapolated probability for click pattern $[1, 1]$ due to the uncertainty of loss values. The fluctuation of the loss value 
is assumed to follow a normal distribution with standard deviation $0.01$. The total number of samples is 500, and the mean and standard deviation of the extrapolated 
probability is 0.2425 and 0.2873, respectively. }
\label{fig:LossFluctuation}
\end{figure}

For the TMSV state example, we consider the effect of probability measurement uncertainty and the fluctuation of loss values, and the results are shown in 
Figs.~\ref{fig:ProbFluctuation} and \ref{fig:LossFluctuation}, respectively. The probability we want to extrapolate is $P_0 ([1,1])= 0.2436$. We choose the photon loss
as $\epsilon = 0.3$ and $\vt{c} = (1.0, 1.3, 1.6, 1.9, 2.2)$, and obtain the extrapolated result $\tilde P ([1,1])= 0.2367$ assuming no fluctuations. 
In the first case, the relative probability measurement uncertainty, as defined in Eq.~\eqref{eq:uncertainty-in-p}, comes from the finite number of collecting tries. 
Here we collect 500 samples of size $N=10^5$. We can see from Fig.~\ref{fig:ProbFluctuation} that the extrapolated probability deviates from $\tilde P ([1,1])$ and the variance is so big such that one sometimes gets negative extrapolated probabilities. However, the peak of the distribution 
is still around 0.2 and the mean value can be calculated to be 0.2373, which is close to the $\tilde P ([1,1])$, and the standard deviation is 0.2531. 
The effect of the loss fluctuation is similar, 
see Fig.~\ref{fig:LossFluctuation}. We assume the loss fluctuation follows a normal distribution with standard deviation 0.01. 
The extrapolated probability also follows a Gaussian-like distribution with big variance, and 
one sometimes gets negative values. However, the peak of the distribution is still around 0.2 and the mean value can be calculated to be 0.2112, 
which is close to the $\tilde P ([1,1])$, and the standard deviation is 0.2873. 
This example shows that the extrapolation amplifies the variance, as indicated by Eq.~\eqref{eq:Variance}. However, it is still practical to get good results by choosing 
an appropriate number of loss values.

\subsection{Nonzero displacement and nonuniform loss}

We have shown that the extrapolation technique works for a GBS device with squeezed vacuum states as inputs. More general input states consist of displacements,
for example in the algorithm to simulate the molecular vibronic spectra in a GBS device \cite{Huh2015}. Here we show that in the presence of displacements the probability can also
be expanded as a series with respect to the loss $\epsilon$, therefore the extrapolation technique still works. We further show that the ``singular" part (the poles) in the probability
expression is independent of the displacements, therefore the improved extrapolation technique also applies. For details see Appendix \ref{ap:displacement}. 

We have studied the simplified case where the photon loss in each mode are the same, which can be considered as a good approximation if 
the linear interferometer is implemented using the Clements' decomposition~\cite{Clements2016}. However, in the realistic implementation 
the photon loss is not uniform, and the situation becomes even worse if the interferometer is implemented using the Reck's decomposition~\cite{Reck1994}. 
We now propose an extrapolation technique for a GBS device with nonuniform loss. 

A universal $M$-mode linear interferometer consists of $N = M(M-1)/2$ beam splitters. Assume that each beam splitter is lossy and is characterized by two lossy 
channels. We order the $N$ beam splitters and label them using integers $k$, and the loss parameters of the corresponding two lossy channels as $\epsilon_{ka}$ and
$\epsilon_{kb}$.  
The probability of detecting a click pattern $\vt{n}$ is a function of all loss parameters, and is denoted as $P(\vt{n}; \vt{\epsilon})$
with $\vt{\epsilon} = (\vt{\epsilon}_a, \vt{\epsilon}_b)$ and $\vt{\epsilon}_a = (\epsilon_{1a}, \epsilon_{2a}, \cdots, \epsilon_{N a})$, $\vt{\epsilon}_b = (\epsilon_{1b}, \epsilon_{2b}, \cdots, \epsilon_{N b})$. 
Define $P^{(ia)}(\vt{n})$ ($P^{(ib)}(\vt{n})$) as the probability by changing only one loss parameter from $\epsilon_{ia}$ ($\epsilon_{ib}$) 
to $c_{ia} \epsilon_{ia}$ ($c_{ib} \epsilon_{ib}$), with $c_{ia}$ and $c_{ib}$ greater than one. 
We then obtain an approximation of $P_0(\vt{n})$ to the second order of the photon loss,
\begin{align}\label{eq:EstimateProb}
\tilde P(\vt{n};  \vt{\epsilon}) = P(\vt{n}; \vt{\epsilon}) - \sum_\mu \sum_{k=1}^N \frac{P^{(k\mu)}(\vt{n}) - P(\vt{n}; \vt{\epsilon})}{c_{k\mu} - 1},
\end{align}
where $\mu = \{ a, b \}$. 

In the experiment, one first measures $P(\vt{n}; \vt{\epsilon})$ with loss $\vt{\epsilon}_a$ and $\vt{\epsilon}_b$; 
then change one of the loss parameters while keeping other parameters unchanged to measure $P^{(ia)}(\vt{n})$ and $P^{(ib)}(\vt{n})$, 
which requires $M(M-1)$ repeats of experiments. 
By combining all these measurement results we get a better approximation of the probability $P_0(\vt{n})$ via Eq.~\eqref{eq:EstimateProb}. This method is advantageous for a large
circuit when the photon loss of each beam splitter is small. To precisely control the loss of each beam splitter could be challenging for an interferometer integrated on a chip, but is
realizable for architectures based on delay loops.

\section{Loss Cancellation}\label{sec:loss-cancellation}

We now discuss another scheme to mitigate the effect of photon loss in a GBS device. It is specifically tailored for photon number detection and is particularly 
suitable for platforms where photon number detection is available. 
One of the advantages of this scheme is that it requires no hardware modifications of the GBS device. The only computational cost
is to calculate a set of coefficients, which we will discuss in details.

\subsection{The general procedure}

\begin{figure}
\includegraphics[width=0.9\columnwidth]{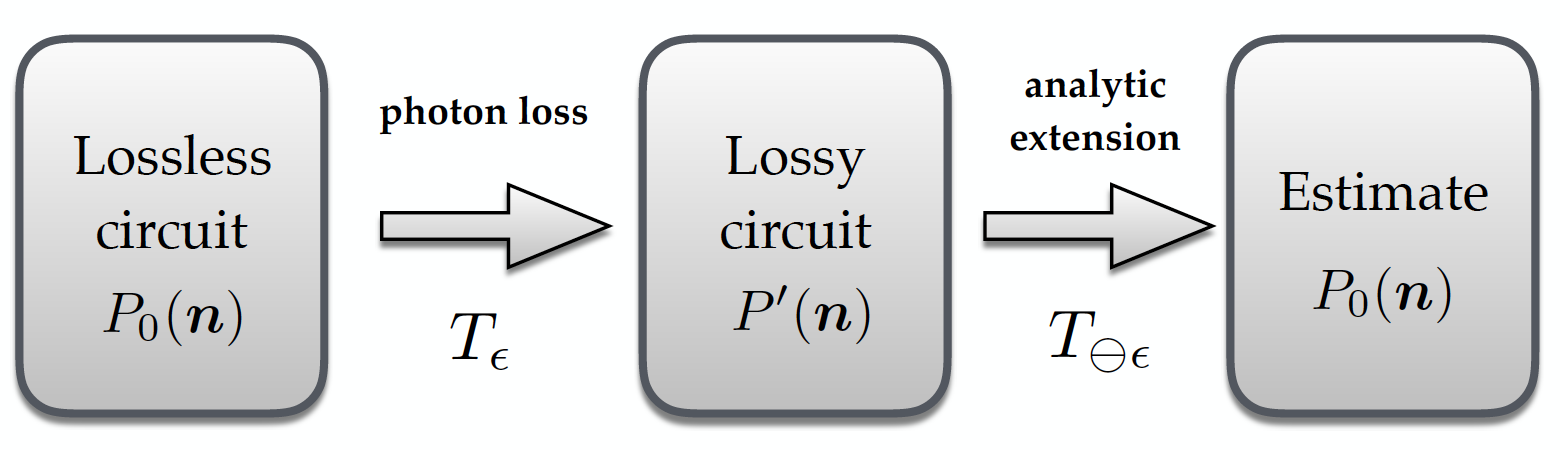}
\caption{ Procedure to cancel photon loss. In the first step, the probabilities of a lossy device are calculated using the probabilities of a loss-free device. This corresponds
to a physical process and is accomplished by applying the operator $T_{\epsilon}$. In the second step, the probabilities of a loss-free device are inferred using 
the probabilities of a lossy device. This is accomplished by applying the operator $ T_{\ominus \epsilon}$ but does not correspond to a physical process. }
\label{fig:robertEM}
\end{figure}

Consider a noisy $M$-mode device with same photon loss in each mode, characterized by a single parameter $\epsilon$. Equivalently, this noisy GBS device can be 
modelled as placing $M$ beam splitters, with the same transmissivity $\eta = 1-\epsilon$, after a lossless GBS device. 
The beam splitters take a Fock state $\vt{n}$ (corresponding to an  $M$-tuple of nonnegative integers $n_j$) to ${\vt n}' \le \vt n$ ($n'_j \le n_j$ for all $j$) with conditional probabilities

\begin{equation}\label{eq:conditionalProb}
P_{\epsilon}(\vt{n}' | \vt{n}) = \prod_{j=1}^M {n_j \choose n'_j} \epsilon^{n_j - n'_j} (1-\epsilon)^{n'_j}.
\end{equation}
Then given probabilities for the  states in the lossless GBS device, we get probabilities for the states in a lossy GBS device.  
This can be expressed in terms of a linear operator $T_{\epsilon}$ on  measures on $\mathbb N^M$.
Define $P_0(\vt{n})$
as the probabilities for Fock states $\vt n$ in the lossless GBS device, then the probabilities in the lossy GBS device are
\begin{align}\label{eq:LossToLossless}
P' (\vt{n}'; \epsilon) = T_{ \epsilon} (P_0)({\vt n}') = \sum_{{\vt n} \ge {\vt n}'} P_{ \epsilon}({\vt n}'|{\vt n}) P_0({\vt n}). 
\end{align}
This involves an infinite series, but convergence is clear (for $0 < \epsilon < 1$) since $\sum_{\vt n} P_0({\vt n}) = 1$ 
and $0 \le P_{\epsilon}({\vt n}'|{\vt n}) \le 1$.
To compute $T_{ \epsilon}(P_0)({\vt{n}'})$  
for a given final Fock state $\vt{n}'$, we do not need to compute conditional probabilities for all $\vt{n}$, but only for those that could 
lead to $\vt{n}'$, i.e., with $\vt{n} \ge \vt{n}' $. 
If $P_0$ has finite support, we only need to include click patterns up to the largest $|\vt{n}|$ (the total number of photons of click pattern $\vt{n}$) for which $P_0(\vt{n}) \ne 0$.  
Even if the support is infinite, in a practical computation we might impose a finite cutoff.

What we have just described, evaluating the probabilities of a lossy GBS device from the probabilities of a loss-free GBS device, is straightforward. However, what we want
is the inverse: using the probabilities of a lossy GBS device to infer the probabilities of a loss-free GBS device. 
This can be accomplished by performing an analytic continuation to the transformation~\eqref{eq:LossToLossless}. 

We first show that the transformation $T_\epsilon$ form a semigroup. 
Suppose we apply two sets of beam splitters to a loss-free
device, with photon losses ${\epsilon}$ and $\mu$, respectively. The two sets of beam splitters can be combined into one set of beam splitters with transmissivity
$(1-\epsilon)(1-\mu)$, namely, with photon losses $\epsilon+\mu-\epsilon \mu$. This implies that
the operators $T_{ \epsilon}$ form a semigroup, with $T_{ \epsilon} T_{ \mu} = T_{{ \epsilon} \oplus  \mu}$, where $ \epsilon \oplus  \mu = \epsilon + \mu - \epsilon \mu$.
Now  the operators $T_{\epsilon}$ can be defined formally for arbitrary complex numbers $\epsilon$, not just those  in $[0,1]$: at least if $P$ is a complex measure of finite support, $T_{\epsilon} (P)$ will be a complex measure with finite support, and $\sum_{\vt{n}} T_{ \epsilon} (P)(\vt{n}) = 1$, and by analytic continuation the formula $T_{ \epsilon} T_{\mu} = T_{\epsilon \oplus \mu}$ is still true.  The formulas are all the same, although the interpretation of Eq.~\eqref{eq:conditionalProb} as a conditional probability is no longer there.

We can use this to go backwards, inferring $P_0$ from $P' = T_{\epsilon} (P_0)$. By setting $\epsilon \oplus \mu = 0$, we find $\mu=\epsilon/(\epsilon-1)$,
which we denote as $\ominus  \epsilon$.
Thus $T_{\ominus  \epsilon} T_{ \epsilon}$ is formally equal to the identity map, so that applying $T_{\ominus  \epsilon}$ formally brings the probabilities
$P' = T_\epsilon (P_0)$ back to the loss-free probabilities $P_0$. This whole procedure is schematically shown in Fig.~\ref{fig:robertEM}. 
In the calculation, one only needs to substitute $\ominus \epsilon = \epsilon/(\epsilon-1)$ for $\epsilon$:

\begin{align}\label{eq:EstimationLC}
{P_0}({\vt{m}})  
= T_{\ominus  \epsilon}(P')({\vt{m}}) = \sum_{\bf n \ge \bf m} P_{\ominus \bf \epsilon}({\vt{m}} \mid {\vt{n}}) P'(\vt{n}) 
\nonumber\\
= \sum_{\vt{n} \ge \vt{m}} \left[ \prod_{j=1}^M {n_j \choose m_j} \left(-\frac{1}{\epsilon}\right)^{m_j} \left(\frac{\epsilon}{\epsilon-1} \right)^{n_j} \right]
 P' (\vt{n}). \nonumber \\
\end{align}
We call this procedure loss cancellation.

Our formal  result Eq.~\eqref{eq:EstimationLC}  involves an infinite series. The convergence of this infinite series 
is not guaranteed in the whole parameter regime. Of course in practice we can only consider finitely many terms, so we might impose a cutoff, but 
unless the series converges, the result of this computation might not be a good approximation to the loss-free probabilities.
See Appendix~\ref{ap:convergenceproof} for a proof that under appropriate conditions, for sufficiently low photon loss the series converges and gives good approximation
to the loss-free probabilities. In an actual experiment, the overall photon loss $\epsilon$ has to be determined first. This can be done by injecting coherent
lights into the circuit and measuring the output coherent lights~\cite{Rahimi-Keshari:13}. 
After the overall photon loss $\epsilon$ being determined, one runs the experiment many times and collects enough data to
estimate the probability $P' (\vt{n})$. Finally, the probability without photon loss ${P_0}({\vt{m}})$ is calculated using Eq.~\eqref{eq:EstimationLC}.

\subsection{Test for a two-mode squeezed vacuum }

The two-mode squeezed vacuum state is given by Eq.~\eqref{eq:TMSV}, from which it is clear that the probabilities for click patterns 
$[i, i]$ are given by
\begin{eqnarray}
P_0([i,i]) = \chi^{2i} (1-\chi^2),
\end{eqnarray}
and zero for other patterns. 
With uniform loss $\epsilon$ we have $P' = T_\epsilon(P_0)$, where for $i \le j$,
\begin{eqnarray*}
&&P'([i,j]; \epsilon) 
\nonumber\\
&=& \big(1-\chi^2 \big) \sum_{k=j}^\infty {k \choose i}{k \choose j} \chi^{2k} \epsilon^{2k-i-j}  (1-\epsilon)^{i+j} \nonumber\\
&=& {j \choose i} \chi^{2j} \epsilon^{j-i} (1-\chi^2) (1-\epsilon)^{i+j} 
\nonumber\\
&& \times {}_2F_{1}(j+1,j+1;\ j-i+1; (\epsilon \chi)^2), 
\end{eqnarray*}
and by symmetry we have $P'([j,i]; \epsilon) = P'([i,j]; \epsilon)$.  
Since $0 < \epsilon < 1$ and $0 < \chi < 1$, the series converges and can be written using a hypergeometric function. 

We now apply the loss cancellation procedure to estimate the probability $P_0([1,1])$. Two cases are considered: in one case we choose the cutoff photon number 
as $n_{\rm max} = 7$, which means only click patterns $[n_1, n_2]$ with $n_1 \ge 1, n_2 \ge 1$ and $n_1 + n_2 \le 7$ are considered,
and the resulting approximation is denoted as $\tilde{P}_7([1, 1])$; while in the other case we choose the cutoff photon number as $n_{\rm max} = 10$ 
and the resulting approximation is denoted as $\tilde{P}_{10}([1, 1])$. We find that the Maclaurin series of $\tilde{P}_7([1, 1])$ in $\epsilon$ begins 
\begin{eqnarray}
\chi^2(1-\chi^2 ) - 16 \chi^8(1-\chi^2 ) \epsilon^6 + O(\epsilon^8),
\end{eqnarray}
and that for $\tilde{P}_{10}([1, 1])$ begins
\begin{align}
 \chi^2(1-\chi^2) + 324 (\chi^{12}-\chi^{14}) \epsilon^{10} + O(\epsilon^{12}),
 \end{align}
where $\chi^2 (1-\chi^2)$  is the value of the actual $P_0([1,1])$. Because the next nonzero coefficients are 
$\epsilon^6$ (for $n_{\rm max} = 7$) and $\epsilon^{10}$ (for $n_{\rm max} = 10$) respectively, 
this should give very good results when $\epsilon$ is small. 

\begin{table*}[ht]
\caption{Approximate the probability $P_0([1, 1])$ using the loss cancellation procedure for cutoff  photon number $n_{\rm max} = 7$ and $n_{\rm max} = 10$. 
The first column lists the loss values and the first row gives the lossless probability $P_0([1, 1])$. 
The second column gives the estimated probability $\tilde{P}_7([1, 1])$ for $n_{\rm max} = 7$, with two subcolumns 
corresponding to input squeezing $r=1/2$ and $r=1$, respectively. The third column gives the estimated probability $\tilde{P}_{10}([1, 1])$ for $n_{\rm max} = 10$.} 
\centering 
\resizebox{0.9\textwidth}{!}{\begin{minipage}{\textwidth}
\centering
\begin{tabular}{@{}c|cc|cc}
\hline \hline
 ~~~~$\epsilon$~~~~ & $n_{\rm max} = 7$ & & $n_{\rm max} = 10$ & \\
    & ~~~$r=1/2$~~~ & ~~~$r = 1$~~~ & ~~~$r = 1/2$~~~ & ~~~$r=1$~~~ \\
    \hline
0.0 & 0.167948 & 0.243596 & 0.167948 & 0.243596\\
0.1 & 0.167948 & 0.243595 & 0.167948 & 0.243596 \\
0.2 & 0.167946 & 0.243502 & 0.167948 & 0.243597\\
0.3 & 0.167914 & 0.241527 & 0.167948 & 0.243697\\
0.4 & 0.167678 & 0.218252 & 0.167953 & 0.247736 \\
0.5 & 0.166384 & 0.008163 & 0.168027 & 0.351743\\
0.6 & 0.160535 & -1.698578 & 0.168753 & 2.555229\\
0.7 & 0.137057 & -15.634539 & 0.174541 & 47.943868 \\
0.8 & 0.049440 & -142.109725 & 0.215083 & 1100.091815\\
\hline \hline
\end{tabular}
\label{tab:LossCancellation}
\end{minipage}}
\end{table*}

Table \ref{tab:LossCancellation} shows the estimated result for $P_0([1, 1])$ using loss cancellation for different input squeezing and cutoff photon number.
For $r =0.5$ and $n_{\rm max} = 7$, the loss cancellation gives very good approximations for photon loss up to $\epsilon =0.5$, and the estimation fails for high
photon loss like $\epsilon > 0.6$. Increasing the cutoff photon number to $n_{\rm max} = 10$ slightly improves the result and pushes the boundary to 
about $\epsilon = 0.6$, but still fails for photon loss $\epsilon > 0.7$. For $r =1.0$ and $n_{\rm max} = 7$, good approximation is achieved for photon loss $\epsilon \le 0.3$
and the result becomes meaningless for $\epsilon \ge 0.5$. 
We discuss this result in detail in Appendix \ref{ap:convergenceproof}.  For $r = 0.5$, our series should converge for all $\epsilon < 1$,
so for any $\epsilon$ we should be able to attain good results by taking a sufficiently large cutoff.  However, for $r = 1.0$ our convergence result only works for $\epsilon < 1/(2\chi) \approx 0.6565$, and for $\epsilon$ greater than that an increased cutoff would be useless (see Appendix \ref{ap:convergenceproof} for more details).

\subsection{Using empirical data}

\begin{table*}[ht]
\caption{Approximation of the probability $P_0([1, 1])$ using empirical data.  
The first column lists the loss values. The second column (``{\bf No.1"}) gives results of the approximation to ${P}_0([1, 1])$ using the normal loss cancellation, with two subcolumns 
corresponding to the mean and standard deviation, respectively. The third column ( ``{\bf No.2}") gives results using the series expansion introduced in Eq.~\eqref{eq:LCseries}.
The fourth column ( ``{\bf No.3}") gives results by taking into account the form of the probability expression in Eq.~\eqref{eq:probpoles}. We choose the input squeezing 
as $r=1/2$ and collect 100 samples of size $10^5$. } 
\centering
\resizebox{0.85\textwidth}{!}{\begin{minipage}{\textwidth}
\centering
\begin{tabular}{@{}c|cc|cc|cc}
\hline \hline
 ~~~~$\epsilon$~~~~ &  {\bf No.1} &  &  {\bf No.2} &  &  {\bf No.3} & \\
    & ~~~mean~~~ & standard deviation & ~~~mean~~~ & standard deviation & ~~~mean~~~ & standard deviation \\
    \hline
0.2 &  0.167905 & 0.001430 & 0.167823 & 0.001562 & 0.168155 & 0.001387 \\
0.5 & 0.166660 & 0.006316 &  0.167998 & 0.004721 & 0.167166 & 0.005806 \\
0.6 & 0.158845 & 0.021262 & 0.170362 & 0.011843 & 0.167394 & 0.011189\\
0.7 & 0.224839 & 0.065665 & 0.189705 & 0.034023 & 0.164793 & 0.034124 \\
0.8 & -0.015223 & 0.199697 & 0.249027 & 0.116481 & 0.166464 & 0.085374 \\
\hline \hline
\end{tabular}
\label{tab:LCempiricaldata}
\end{minipage}}
\end{table*}

In a typical application, we will use an empirical distribution $\widehat{P}$ from measured data as an approximation to $P'$, and compute
our approximation to $P_0({\vt{n}})$ as $T_{\mu}(\widehat{P})(\vt{n})$. Suppose the empirical data is obtained from a sample of size $N$, 
the estimator $\widehat{P}(\vt{n})$ has mean $P' (\vt{n})$,
variance $P' (\vt{n}) [1 -P' (\vt{n})]/N$, and covariances 
\begin{eqnarray}
\text{Cov} \big( \widehat{P}(\vt{n}), \widehat{P}(\vt{n}') \big)= - P' (\vt{n}) P' (\vt{n}')/N
\end{eqnarray}
for $\vt{n} \ne \vt{n}'$. If we write $$T_{\mu} (\widehat{P})(\vt{n}') = \sum_{\vt{n}} P_\mu(\vt{n}'|\vt{n})  \widehat{P}(\vt{n}),$$
then this has mean $T_{\mu} (P') (\vt{n}') = P_0(\vt{n}')$ and variance
\begin{eqnarray}\label{eq:LCvariance}
&&\frac{\sum_{\vt{n}} P_\mu(\vt{n}'|\vt{n}) ^2 P' (\vt{n})}{N}-\frac{ \big[ \sum_{\vt{n}} P_\mu(\vt{n}'|\vt{n}) P' (\vt{n}) \big]^2}{N}  \nonumber\\
&=&  \frac{\sum_{\vt{n}} P_\mu(\vt{n}'|\vt{n}) ^2 P' (\vt{n})}{N} - \frac{P_0(\vt{n}')^2}{N}. 
\end{eqnarray}

It should be noted that even when the infinite series for $T_{\mu} (P') (\vt{n})$ converges, the variance might not.
Practically speaking, this means that ``outliers" could have a large influence on the variance. Rather than use all the click patterns that appear in our sample, it may be better to
impose a fixed cutoff.  This means our estimator will no longer be unbiased, but it may be more stable.

For several different values of $\epsilon$ with $\chi = \tanh(1/2)$, we took $100$ samples of size $10^5$ from the distribution $P_0$, added loss $\epsilon$ by letting each photon 
survive or disappears with probabilities $1-\epsilon$ and $\epsilon$, and then took the estimator $T_{\ominus \epsilon} (\widehat{P})([1,1])$.  
Recall that the correct value is  $0.167948$. The results are shown in Table~\ref{tab:LCempiricaldata}.

As a function of $\mu$, $T_\mu(P)(\vt{n})$ is an analytic function
which we can expand in a power series around $\mu = 0$, the coefficients
involving probabilities of various click patterns. The terms involving higher powers of $\mu$ will contain click patterns with more extra photons, which will have low probability of being observed but may have large coefficients. In a simulation, a few of these ``outlier" click patterns will often be observed, and  
 this can have a bad effect on the accuracy of our estimate.  It can be better to
only use a limited number of coefficients.  The estimator will no longer be unbiased,
but the variance may decrease significantly.
To illustrate this we consider the example of a TMSV state as before. To fourth order of $\mu$, the series for $T_\mu (P)([1,1])$ reads
\begin{widetext}
\begin{eqnarray}\label{eq:LCseries}
&&T_\mu (P)([1,1]) 
\nonumber\\
&=&  P([1,1]) + \big\{-2 P([1,1]) + 2 P([2,1]) + 2 P([1,2]) \big\} \mu
+ \big\{ P ( [1,1])-4 P ( [2,1] ) -4 P ( [1,2] )
\nonumber\\
&&
 +3 P ( [3,1] ) + 4 P( [2,2] ) +3 P ( [1,3] ) \big\} \mu^2 + \big\{ 2 P ( [2,1] ) 
+2 P ( [1,2] ) - 6 P ( [3,1] )  -8 P ( [2,2] ) 
\nonumber\\
&&
-6 P ( [1,3] ) + 4 P ( [4,1] ) +6 P ( [3,2] ) +6 P ( [2,3] ) +4 P ( [1,4] ) \big\} \mu^3
+ \big\{ 3 P( [3,1]) +4 P ( [2,2] ) 
\nonumber\\
&&
+3 P ( [1,3] ) -8 P ( [4,1] ) -12 P( [3,2] ) -12 P ( [2,3] ) -8 P ( [1,4] ) +5 P ( [5,1] ) 
+8 P( [4,2] ) 
\nonumber\\
&&
+ 9 P( [3,3] )  +8 P( [2,4] ) +5 P ( [1,5] ) \big\} \mu^4. 
\end{eqnarray}
\end{widetext}
Now by replacing $\mu$ and $P$ in Eq.~\eqref{eq:LCseries} by $\ominus \epsilon$ and $\widehat P$, respectively, we obtain 
an estimator $T_{\ominus \epsilon} (\widehat{P})([1,1])$ for $P_0([1, 1])$.  With $r = 1/2$ and 
$100$ samples of size $10^5$, we get results with a significant improvement, see Table~\ref{tab:LCempiricaldata}. 

Further improvements are possible if we take advantage of knowledge of the form of $T_{\nu}(P_0)$.  In
the two-mode example we know (see Sec.~\ref{Sec:SecondExtrapolation}) that $T_{\nu} (P_0)([1,1])$ should have the form 
\begin{eqnarray}\label{eq:probpoles}
T_\nu (P_0)([1,1]) = \frac{A(\nu)}{(1-\nu^2 \chi^2)^3},
\end{eqnarray}
where $A(\nu)$ is a polynomial, and in fact we know $A(\nu)$ has degree $\le 4$.
Note that $\nu = \epsilon \oplus \mu = \epsilon + \mu - \epsilon \mu$ so $\mu = (\nu - \epsilon)/(1-\epsilon)$.
\begin{eqnarray}
A(\nu) &=& (1 - \nu^2 \chi^2)^3 \, T_\nu (P_0)([1,1]) 
\nonumber\\
&=&
 (1-\nu^2 \chi^2)^3 \, T_\mu (P)([1,1]),
\end{eqnarray}
where $P = T_\epsilon (P_0)$.  If we expand the right hand side in a power series in $\nu-\epsilon$, since the left hand side
is a polynomial of degree $\le 4$ the terms in higher powers on the right should be $0$.  We can take that series to order $4$,
evaluate at $\nu=0$ using the empirical distribution $\widehat{P}$ instead of $P$, and the result should be a 
good approximation of $A(0) = P_0([1,1])$. We tried this for $r = 1/2$ with $100$ samples of size $10^5$, with quite good results, see Table~\ref{tab:LCempiricaldata}.  
For $\epsilon = 0.9$, the mean and standard deviation are 0.159427 and 0.368437, respectively,  which are not wildly off the mark.

\section{Eight-mode example}\label{sec:example}

We have discussed two schemes to mitigate the effect of photon loss and showcased their performance for a two-mode squeezed state. An important question is whether
these error mitigation techniques can be applied to a large GBS device, which is more relevant to practical applications. 
When the circuit size increases, an immediate issue arises as that the 
size of the Hilbert space for a fixed total photon number increases, so the probability of detecting a single click pattern decreases. It is not practical to estimate a tiny probability
using a GBS device. It is thus necessary to consider the probability of a collection of click patterns, the coarse grained probability. 
One of the useful coarse grained probabilities is the orbit probability~\cite{bradler2018graph}, which is critical in solving the graph isomorphism and graph similarity problems. 
An orbit $O_{\vt{n}}$ is defined as a collection of click patterns including all permutations of the click pattern $\vt{n}$. The orbit probability $P_{O_{\vt{n}}}$
is the sum of all click-pattern probabilities inside an orbit $O_{\vt{n}}$. Although we introduce the mitigation schemes by estimating the probability of a single click pattern,
we show here that they can also be used to estimate the orbit probability, as well as other coarse grained probabilities. 

We consider a book graph of size eight, with adjacency matrix 
\begin{eqnarray}
A = 
	\begin{pmatrix}
	0 & 1 & 1 & 0 & 1 & 0 & 1 & 0 \\
	1 & 0 & 0 & 1 & 0 & 1 & 0 & 1 \\
	1 & 0 & 0 & 1 & 0 & 0 & 0 & 0 \\
	0 & 1 & 1 & 0 & 0 & 0 & 0 & 0 \\
	1 & 0 & 0 & 0 & 0 & 1 & 0 & 0 \\
	0 & 1 & 0 & 0 & 1 & 0 & 0 & 0 \\
	1 & 0 & 0 & 0 & 0 & 0 & 0 & 1 \\
	0 & 1 & 0 & 0 & 0 & 0 & 1 & 0 \\
	\end{pmatrix}. 
\end{eqnarray}
The adjacency matrix $A$ has to be doubled and rescaled to be $cA \oplus A$, so that it can be encoded into an 8-mode pure Gaussian state $\sigma_0$~\cite{bradler2018graph}, 
which is generated by injecting eight pure single-mode squeezed vacuum states into a loss-free interferometer. Here we choose 
$c = 0.25$ so that the maximum input squeezing is about $7.25$ dB, which is accessible for current quantum optics experiments. When the circuit is lossy, the generated 
state is different from $\sigma_0$. We now use the proposed error mitigation schemes to estimate the orbit probabilities for a loss-free circuit. 

\begin{figure*}[ht]
\centering
		\subfigure[]{\includegraphics[width=0.35\textwidth]{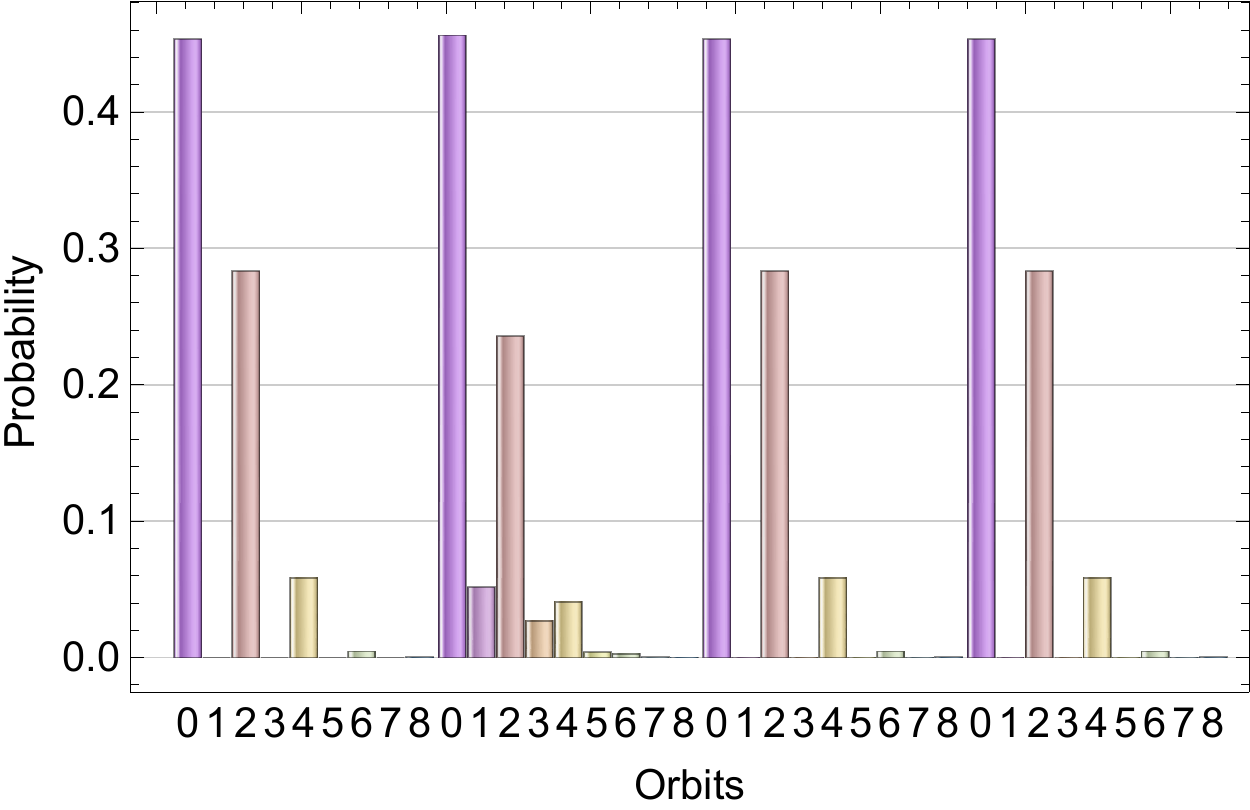}}\qquad \qquad
		\subfigure[]{\includegraphics[width=0.35\textwidth]{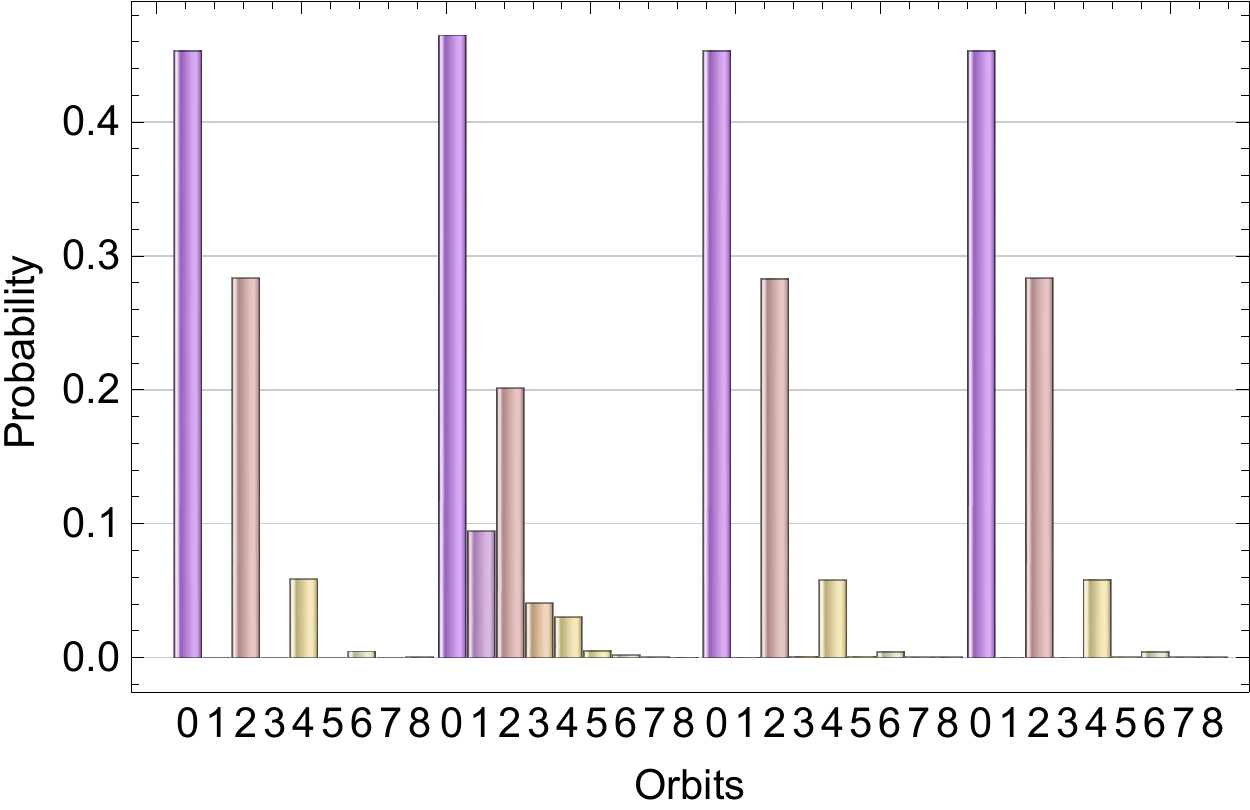}}\\
		\subfigure[]{\includegraphics[width=0.35\textwidth]{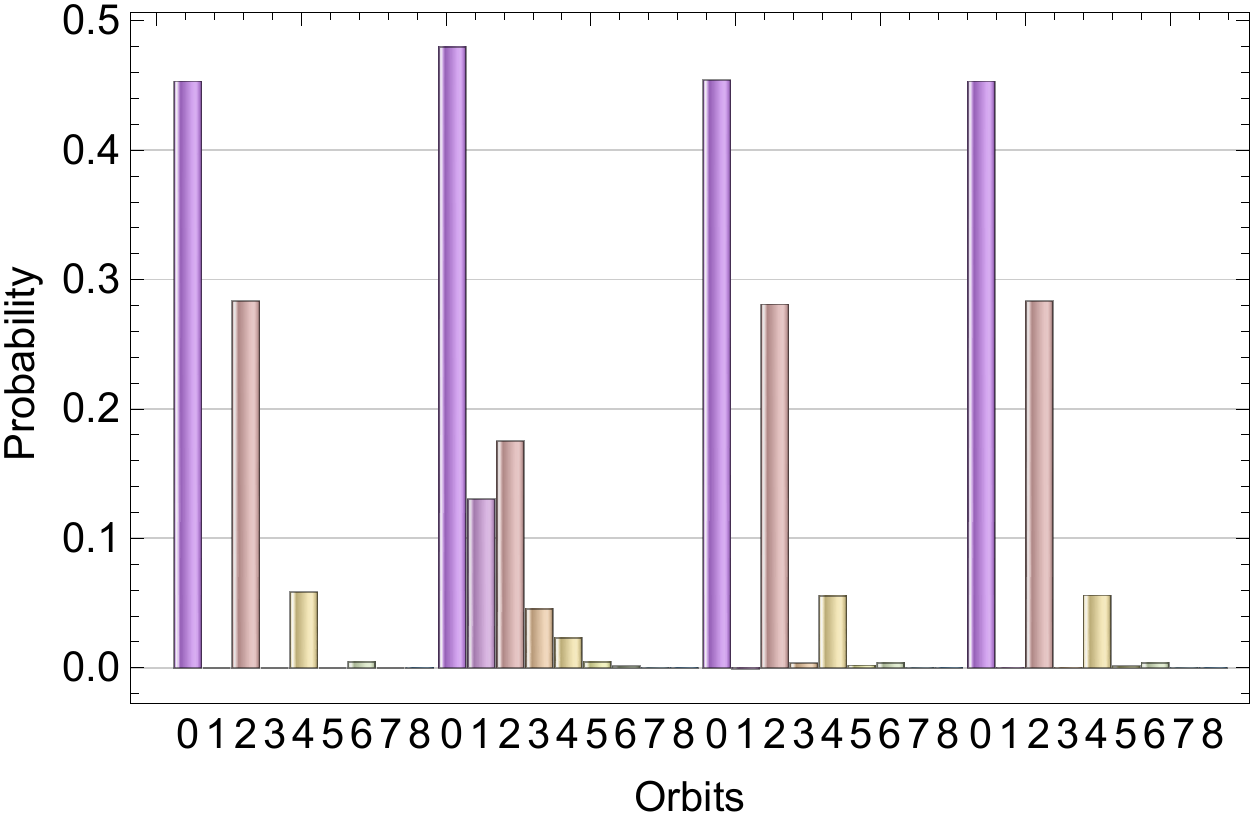}}\qquad \qquad
		\subfigure[]{\includegraphics[width=0.35\textwidth]{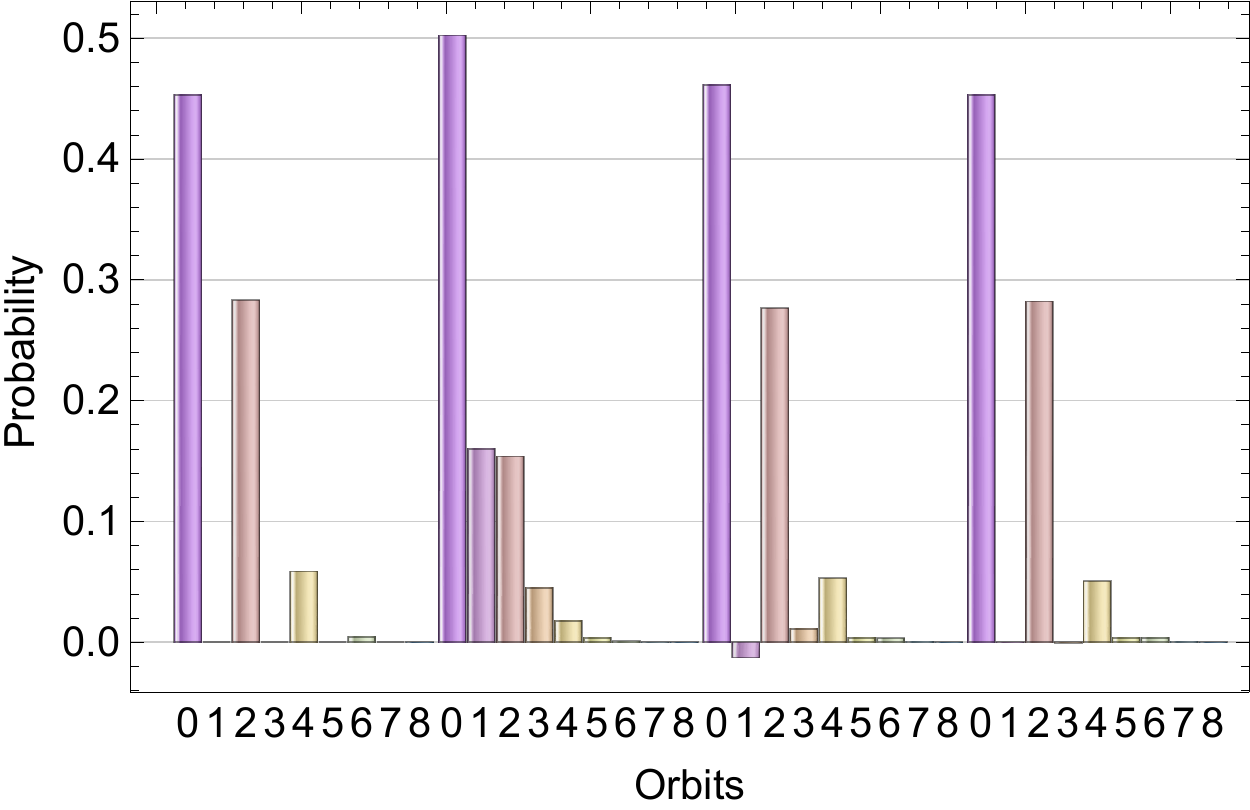}}
	\caption{Estimate orbit probabilities using extrapolation technique. In each subfigure there are four histograms, they represent the orbit probabilities without photon loss, with photon loss, with extrapolation and with the improved extrapolation, respectively. The loss values for (a), (b),  (c) and (d) are $\epsilon = 0.1, 0.2, 0.3$ and $0.4$, respectively. }
	\label{fig:p1111}
\end{figure*}

\begin{widetext}

\begin{table}
\caption{Mitigate photon loss for the orbit probability $P_0([00001111])$. The first column lists the loss values and the second column gives probabilities without doing any 
error mitigation. The third and fourth columns show probabilities after performing extrapolation and the improved extrapolation, respectively. The last column lists probabilities 
with the loss cancellation scheme.  } 
\centering 
\resizebox{0.9\textwidth}{!}{\begin{minipage}{\textwidth}
\centering
\begin{tabular}{ c | c | c | c | c } 
\hline\hline 
{ \bf Photon loss}  & {\bf No mitigation} & {\bf Extrapolation} & {\bf Improved extrapolation} & {\bf Loss cancellation} \\
\hline 
$\epsilon = 0.0 $ & 0.058419 & 0.058419 & 0.058419 & 0.058419 \\ 
$\epsilon = 0.1$ & 0.040659 & 0.058371 & 0.058406 & 0.058582 \\ 
$\epsilon = 0.2$ & 0.030128 & 0.057676  & 0.058019 & 0.058756 \\
$\epsilon = 0.3$ & 0.023141 & 0.055638 & 0.055843 & 0.056712 \\ 
$\epsilon = 0.4$ & 0.017815 &  0.053008 & 0.050619 & 0.040986  \\
$\epsilon = 0.5$ & 0.013216 &  0.052349 & 0.045251 & $-0.042469$ \\
$\epsilon = 0.6$ & 0.008965 & 0.039900 & 0.050386 & $-0.446576$ \\
$\epsilon = 0.7$ &0.005086  &  $-0.189112$ & 0.081490 & $-2.413260$ \\
\hline 
\end{tabular}
\label{tab:compare}
\end{minipage}}
\end{table}

\end{widetext}

Consider orbits with at most one photon in each mode up to 8 total photons: 
vacuum, $[1]$, $[11]$, $[111]$, $[1111]$, $[11111]$, $[111111]$, $[1111111]$ and $[11111111]$, 
where we have omitted ``0" in the click pattern to simplify the notation and used a single click pattern to represent an orbit. The orbit probabilities in the absence of photon loss are
$(0.453, 0., 0.283, 0., 0.058, 0., 0.0044, 0., 0.00011)$. In the presence of photon loss, the photons tend to populate toward lower photon number orbits. We now apply
the extrapolation technique to estimate the loss-free orbit probabilities. To perform the extrapolation, we choose five loss values and the vector $\vt{c}$ is chosen as
$\vt{c} = (1.0, 1.1, 1.2, 1.3, 1.4)$. 
The results are shown in Fig.~\ref{fig:p1111}. We can see that for low photon loss, both the extrapolation and the improved extrapolation
work very well, giving good approximations for most orbit probabilities. When the photon loss increases, some estimated orbit probabilities from extrapolation lost accuracy, e.g., the 
probability of the orbit with one photon becomes negative. However, for the improved extrapolation with poles removed we still obtain a good approximation, showing 
its advantage for high photon loss.  
We also use the loss cancellation method to estimate the orbit probability $P_0([00001111])$.  With cutoff photon number $n_{\rm max} = 7$, we find that good approximation
can be obtained for photon loss $\epsilon \le 0.3$, see Table~\ref{tab:compare}. This demonstrates that the proposed error mitigation techniques still work for a bigger 
circuit for reasonable amount of photon loss.

\section{Summary}\label{sec:conclusion}

We have proposed two schemes to mitigate the effect of photon loss in a GBS device. The first scheme is based on the extrapolation technique 
and requires a small modification of the GBS circuit: to increase the photon loss of the circuit. The second scheme requires no modifications of the
circuit and thus is hardware efficient. One only needs to measure the probabilities of a lossy circuit and then linearly combine them in an appropriate way. The computational
cost is to calculate the linear combination coefficients. We tested these error mitigation techniques in a two-mode and an eight-mode GBS circuits, and 
showed that they work extremely well for low photon loss, and also give fairly good approximations for relatively high photon loss. 

In realistic experiments, the accuracy of the measured probabilities is limited by the finite number of samples and experimental imperfections. We show that 
the extrapolated probability is sensitive to the measured probabilities, and sometimes one gets apparently meaningless results like negative probabilities. This 
can be overcome by performing multiple experiments and taking the mean value as the estimate for the loss-free probability. The requirement of multiple experiments
should be considered as the classical computational cost. 

While the procedure that we have described so far is focused on GBS devices, similar ideas can also be applied to other near-term photonic architectures such as Boson Sampling.
Boson Sampling differs from GBS in that expressions for click-pattern probabilities are related to permanents of transformation matrices rather than to Hafnians as is the case in GBS.
Devising a concrete procedure that accounts for this difference and mitigated Boson Sampling probabilities is an open problem.
Another important research direction is to devise methods for error mitigation in sampling problems, which may require recovering the quantum states instead of recovering the
expectation value of observables.


{\bf Acknowledgement}: We thank Mark Wilde, Seth Lloyd, Kang Tan, Dylan Mahler for insightful discussions.

\bibliographystyle{plain}

\onecolumn 
\appendix

\section{Uniform loss approximation}\label{ap:ULA}

In this appendix, we briefly review the physical implementation of a linear optics interferometer and how to model the photon loss, and
demonstrate under what conditions the uniform loss approximation is valid. 
A reconfigurable linear optics interferometer consists of an array of tunable beam splitters and phase shifters. Practically, a tunable beam splitter is implemented by
two static 50:50 beam splitters and two phase shifters. The transmission and reflection coefficients of the tunable beam splitter can be varied by changing the phases of the phase 
shifters~\cite{Reck1994}. Due to the imperfect implementation of the tunable beam splitter, photons may be lost when going through it. 
Here we assume that the loss rate of each of the two modes are the same. 
Mathematically, the lossy tunable beam splitter can be modelled by adding two beam splitters with reflection coefficient the same as the loss rate after a perfect
tunable beam splitter. The photons may also be lost when travelling through the medium between adjacent tunable beam splitters, 
e.g., the fibre or waveguide. Similarly, this lossy channel can be modelled by a beam 
splitter, which can be effectively combined with the beam splitter that models the loss of the tunable beam splitter. 

Any linear unitary matrix $U(N)$ can be decomposed into a product of a sequence of $2 \times 2$ unitary matrices, 
which correspond physically to the tunable beam splitters~\cite{Reck1994}. There are two main schemes to perform the decomposition:
the Reck's scheme~\cite{Reck1994} and the Clements' scheme~\cite{Clements2016}. The former implements a linear unitary transformation by arranging the 
tunable beam splitters in a triangular configuration. The photon entering different mode will experience very different path length. Given that the loss rate of each tunable 
beam splitter is almost the same, the overall loss rate for each mode is quite different, resulting in nonuniform loss. The Clements' scheme implements a linear unitary
transformation by arranging the tunable beam splitters in a rectangular configuration, in which the photon entering different mode will experience almost
the same path length. If the loss rate of each tunable beam splitter is almost the same, then the overall loss rate for each mode is almost the same, resulting in uniform loss. 

Consider an $M$-mode linear interferometer. If the interferometer is perfect, then its input-output relation is given by
$\hat{\vt{b}} = U \hat{\vt{a}}$,
where $\hat{\vt{a}} = (\hat{a}_1, \hat{a}_2, \cdots, \hat{a}_M)$ with $\hat{a}_i$ the input annihilation operator, 
$\hat{\vt{b}} = (\hat{b}_1, \hat{b}_2, \cdots, \hat{b}_M)$ with $\hat{b}_i$ the output annihilation operator, and $U$ is an $M \times M$ unitary matrix representing the transformation
of the linear interferometer. If the linear interferometer is lossy, then the input-output relation has to be modified as~\cite{Giovannetti2015, GarciaPatron2019simulatingboson}
\begin{eqnarray}\label{eq:LossIF}
\hat{\vt{b}} = A \hat{\vt{a}} + \sqrt{I - A A^\dag} \, \hat{\vt{e}},
\end{eqnarray}
where $\hat{\vt{e}}$ represents the environmental modes and $A$ is a complex matrix satisfying $A A^\dag \le I$. 
The matrix $A$ can be decomposed as $A = V \hat{\lambda} W$, where $V$ and  $W$ are unitary matrices, and 
$\hat{\lambda} = \text{diag} \{\sqrt{\eta_1}, \sqrt{\eta_2}, \cdots, \sqrt{\eta_M} \}$
with $\eta_i \in [0, 1]$. The singular decomposition of $A$ implies that the transformation of a lossy linear interferometer, Eq.~\eqref{eq:LossIF},
is mathematically equivalent to first applying a unitary transformation $W$, followed by $M$ lossy channels with transmission coefficients $\eta_i$ (or loss values $1 - \eta_i$), 
then applying another unitary transformation $V$. 

In general, the singular eigenvalues of $A$ are different and no further simplification can be made. However, 
when the photon entering each mode goes through almost the same number of tunable beam splitters and experiences almost the same path length, then the singular
eigenvalues $\sqrt{\eta_i}$ are almost the same. This implies the diagonal matrix $\hat{\lambda}$ is close to an identity matrix, $\hat{\lambda} = \sqrt{\eta} \,I$, where 
$\eta$ is the overall transmission coefficient of each mode. In this case the matrix $\hat{\lambda}$ commute with $V$ and the matrix $A$ can be rewritten as 
$A = \hat{\lambda} \tilde{U}$, where $\tilde{U} = V W$ is a unitary transformation. This shows that a linear interferometer with uniform loss is equivalent to a perfect 
linear interferometer followed by $M$ lossy channels with the same transmission coefficient.

\section{Nonzero displacement}\label{ap:displacement}

When the displacements are not zero, the probability of the click pattern $\vt{n}$ is given by~\cite{kruse2019detailed}
\begin{eqnarray}\label{eq:ProbabilityGeneral-displace}
P(\vt{n}) &=& \frac{\exp\big(-\frac{1}{2} \vt{d}^{\dag} \sigma_Q^{-1} \vt{d} \big) }{\vt{n}! \, \sqrt{\text{det} \, \sigma_Q}} \prod_{k=1}^M \bigg( \frac{\partial^2}{\partial \alpha_k \partial \alpha_k^* }\bigg)^{n_k}
\exp\bigg(\frac{1}{2} \vt{\alpha}_v^{\top} A \vt{\alpha}_v + \vt{F}^{\dag} \vt{\alpha}_v \bigg) \bigg|_{\vt{\alpha}_v = \vt{0}},
\end{eqnarray}
where $\vt{F} = \sigma_Q^{-1} \vt{d} $. In the presence of photon loss, the probability $P(\vt{n}; \epsilon)$ is obtained by replacing $\sigma_Q$ and $A$  in 
Eq.~\eqref{eq:ProbabilityGeneral-displace} by $\sigma_Q(\epsilon)$ and $A(\epsilon)$ given by Eqs.~\eqref{eq:sigmaQloss} and \eqref{eq:Aexpansion}, respectively. 
Since both $A(\epsilon)$ and $\sigma_Q(\epsilon)$ have series expansions with respect to $\epsilon$, a similar series expansion for 
$P(\vt{n}; \epsilon)$ like Eq.~\eqref{eq:ProbabilitySeries} can be obtained, thus the extrapolation applies. 

By using the relation between $A$ and $\sigma_Q$ in Eq.~\eqref{eq:Amatrix}, we find $[\sigma_Q(\epsilon)]^{-1} = \mathbb{I}_{2M} - X_{2M} A(\epsilon)$,
and by further using the decomposition of $A(\epsilon)$ in Eq.~\eqref{eq:Aepsilon}, we have
\begin{align}\label{eq:SigQepsilon}
[\sigma_Q(\epsilon)]^{-1} &=
U \bigoplus_{k=1}^{M} 
	\begin{pmatrix}
	 \frac{1-\tanh(r_k)}{1-\epsilon \tanh(r_k)} & 0 \\
	0 &  \frac{1+\tanh(r_k)}{1+\epsilon \tanh(r_k)} 
	\end{pmatrix}
U^{\dag}
\nonumber\\
&=
\frac{1}{\mathcal{P}(\epsilon, \tilde{\vt{r}})} U  \bigoplus_{k=1}^{M} 
	\begin{pmatrix}
	 \frac{1-\tanh(r_k)}{1-\epsilon \tanh(r_k)} \mathcal{P}(\epsilon, \tilde{\vt{r}}) & 0 \\
	0 &  \frac{1+\tanh(r_k)}{1+\epsilon \tanh(r_k)}  \mathcal{P}(\epsilon, \tilde{\vt{r}})
	\end{pmatrix}
U^{\dag}. 
\end{align}
By using Eq.~\eqref{eq:SigQepsilon} we can show that 
\begin{eqnarray}
\prod_{k=1}^M \bigg( \frac{\partial^2}{\partial \alpha_k \partial \alpha_k^* }\bigg)^{n_k}
\exp\bigg(\frac{1}{2} \vt{\alpha}_v^{\top} A(\epsilon) \vt{\alpha}_v + \vt{F}^{\dag} \vt{\alpha}_v \bigg) \bigg|_{\vt{\alpha}_v = \vt{0}}
= \frac{1}{\mathcal{P}^N(\epsilon, \tilde{\vt{r}})} \, \mathbb{P}_d(\epsilon),
\end{eqnarray}
where $\mathbb{P}_d(\epsilon)$ is a polynomial of $\epsilon$. Therefore, the probability of measuring a click pattern $\vt{n}$ in the presence of photon loss and 
displacements can be written as
\begin{eqnarray}\label{eq:Probability-displace}
P(\vt{n}; \epsilon) = \frac{\exp\big\{-\frac{1}{2} \vt{d}^{\dag}[\sigma_Q(\epsilon)]^{-1} \vt{d} \big\} }{\vt{n}!}  \bigg( \prod_{k=1}^M \frac{1}{\cosh r_k} \bigg) 
\frac{1}{\mathcal{Q} (\epsilon, \vt{r}) \mathcal{P}^N(\epsilon, \tilde{\vt{r}})} \, \mathbb{P}_d(\epsilon). 
\end{eqnarray}
By using Eq.~\eqref{eq:SigQepsilon}, the quantity $\vt{d}^{\dag} [\sigma_Q(\epsilon)]^{-1} \vt{d}$ can be simplified as
\begin{align}
\vt{d}^{\dag}[\sigma_Q(\epsilon)]^{-1} \vt{d} &=& 
( \vt{d}^{\dag} U)  \bigoplus_{k=1}^{M}
	\begin{pmatrix}
	 \frac{1-\tanh(r_k)}{1-\epsilon \tanh(r_k)} & 0 \\
	0 &  \frac{1+\tanh(r_k)}{1+\epsilon \tanh(r_k)} 
	\end{pmatrix}
(U^{\dag} \vt{d} )
=
\vt{d}_{\rm in}^{\dag} \bigoplus_{k=1}^{M}
	\begin{pmatrix}
	 \frac{1-\tanh(r_k)}{1-\epsilon \tanh(r_k)} & 0 \\
	0 &  \frac{1+\tanh(r_k)}{1+\epsilon \tanh(r_k)} 
	\end{pmatrix}
\vt{d}_{\rm in},
\nonumber
\end{align}
where $\vt{d}_{\rm in} = U^{\dag} \vt{d}$ is the input displacement vector. Therefore the exponential term 
$\exp\big\{-\frac{1}{2} \vt{d}^{\dag}[\sigma_Q(\epsilon)]^{-1} \vt{d} \big\}$ is known if the input squeezing parameters and displacements are known,
and does not need to be extrapolated. The only term that is unknown is the polynomial $\mathbb{P}_d(\epsilon)$, which requires extrapolation. The 
improved extrapolation technique can be applied here since the poles are determined by $\mathcal{Q} (\epsilon \, \vt{r}) $ and $ \mathcal{P}(\epsilon, \tilde{\vt{r}})$ which 
can be removed by simply moving them to the left hand side of Eq.~\eqref{eq:Probability-displace}.

\section{Convergence of the loss cancellation procedure}\label{ap:convergenceproof}

In the Loss Cancellation procedure,
 $T_{\ominus \bf \epsilon}(P')$ is expressed by an infinite series.  We may ask how small $|\epsilon|$ must be to ensure that the series converges, i.e.\
that our approximations approach a limit as the cutoff goes to $\infty$.  This will depend on decay properties of the distribution $P$.

{\bf Definition} The decay radius of the distribution $P$ is the infimum of $u > 0$ such that there exists a constant $C(u)$ with $|P({\vt{n}})| \le C(u) u^{|{\vt{n}}|}$ for all $\vt{n}$.

{\bf Lemma} {\it Suppose $P$ has decay radius $\le t$, i.e. for any $u \in (t,1) $ there exists a constant $C(u)$  such that $|P({\vt{n}})| \le C(u) u^{|{\vt{ n}}|}$ for all $\vt{n}$.
Then for any $ \epsilon$ (not necessarily positive), the series for $T_\epsilon (P)$ converges if  $|\epsilon|  < 1/t$, and the result has decay radius
 $\le \dfrac{|1-\epsilon| t}{1-|\epsilon| t}$. }

\begin{proof}
We have 
\begin{align*}
\left|T_{\epsilon}(P)(\vt{n}')\right| \le \sum_{k \ge |{\vt{n}'}|} L(\epsilon, k, \vt{n}'),
\end{align*}
where
\begin{align*}
L(\epsilon, k, \vt{n}') &= \sum_{\vt{n}: \vt{n} \ge \vt{n}', |\vt{n}|=k} \left| \text{Prob}(\vt{n}'|\vt{n}) P(\vt{n})\right|\cr
&\le \left( \sum_{\vt{n}: \vt{n} \ge \vt{n}', |\vt{n}|=k} \prod_{j=1}^m {n_j \choose n'_j}\right) |\epsilon|^{k-|\vt{n}'|}  |1-\epsilon|^{|\vt{n}'|} C(u) u^k.
\end{align*}
We claim that 
\begin{align*}
\sum_{\vt{n}: \vt{n} \ge \vt{n}', |\vt{n}|=k} \prod_{j=1}^m {n_j \choose n'_j} = {{k + m-1} \choose {|\vt{n}'|+m-1}}.
\end{align*}
The claim can be proven by a  ``stars and bars'' argument.  Let $|\vt{n}'| = n$.  The left side is the number of objects consisting of an $m$-tuple $\vt{n} \ge \vt{n}'$ and, for each $j \in \{1,\ldots, m\}$,
a subset of cardinality $n'_j$ of $[1,\ldots, n_j]$.  The right side counts subsets of cardinality $n+m-1$ of $[1,\ldots, k+m-1]$.  These subsets can be
placed in one-to-one correspondence with the objects on the left side as follows. If the subset is $[T_1, \ldots, T_{n+m-1}]$, we designate $T_{n'_1+1}, T_{n'_1 + n'_2 + 2}, \ldots, T_{n'_1 + \ldots + n'_{m-1}+m-1}$ as ``bars'' $b_1, \ldots, b_{m-1}$ and the others as ``stars'', so that there are  $n$ stars separated by bars into groups of $n'_1, \ldots, n'_m$.  We take $n_1 = b_1-1$,
$n_2 = b_2 - b_1 - 1, \; \ldots,\; n_m = k+m-1-b_{m-1}$.   The subset of cardinality $n'_j$ of $[1,\ldots, n_j]$ then consists of the $j$'th group of ``stars'' translated to the left (if $j > 1$) by $b_j$.  

Now, with $|\vt{n}'|=n$,  we have 
$$L(\epsilon, k, \vt{n}')  \le C {{k + m - 1} \choose {n + m - 1}} |\epsilon|^{k - n} |1-\epsilon|^n u^k.$$
Since $$  {{k + m - 1} \choose {n + m - 1}} \le \frac{(k+m-1)^{n+m-1}}{(n+m-1)!},$$
the sum over $k$ converges absolutely if $|\epsilon u| < 1$. 

Now since $$ \sum_{k=n}^\infty {k+m-1 \choose n+m-1} z^k = \sum_{j=0}^\infty {j+n+m-1 \choose n+m-1} z^{n+j} = \frac{z^n}{(1-z)^{n+m+2}}$$
for $|z| < 1$, we get a bound
$$ |T_\epsilon (P)(\vt{n}')| \le C \frac{|1-\epsilon|^n u^n}{(1-|\epsilon|u)^{n+m-2}},$$
so that $T_\epsilon(P)$ has decay radius $\le \dfrac{|1-\epsilon| t}{1 - |\epsilon| t}$. 
\end{proof}

Now we apply the lemma to loss cancellation.
Suppose the lossless distribution $P$ has decay radius $t$.  If $0 < \epsilon < 1/t$, the distribution with loss $P' = T_\epsilon(P)$ has decay radius $\le \dfrac{(1-\epsilon)t}{1 - \epsilon t}$.  Then with $\ominus \epsilon = \epsilon/(\epsilon - 1)$, the series for $P = T_{\ominus \epsilon} (P')$ converges if $\left|(\ominus \epsilon) \dfrac{(1-\epsilon)t}{1 - \epsilon t}  
\right| < 1$, and this is equivalent to
$$  \epsilon < \frac{1}{2t}. $$

The two-mode example is exceptional in that the decay radius can easily be seen to be  $\chi = \tanh(r)$. so we want $\epsilon < 1/(2\chi)$ to ensure convergence.
For squeezing $r=1/2$, $1/(2\chi) > 1.08$, so the series converges for all $\epsilon \in (0,1)$.
But for squeezing $r = 1$, $1/(2 \chi) \approx 0.6565$.

In general it may be difficult to predict in advance the decay radius for $P$, but we may conjecture that it is typically finite and nonzero.  The loss-cancellation procedure can then be
expected to work very well if $\epsilon$ is sufficiently small, but very poorly when $\epsilon$ is too large.

\end{document}